\documentclass[12pt]{iopart}
\usepackage{setstack}
\usepackage{graphicx}
\usepackage{iopams}
\usepackage{color}

\begin{document}

\title{Second-law-like inequalities with information and their interpretations}

\author{Jordan M. Horowitz$^1$ and Henrik Sandberg$^2$}
\address{$^1$ Department of Physics, University of Massachusetts at Boston, Boston, MA 02125, USA}
\address{$^2$ Department of Automatic Control, KTH Royal Institute of Technology, Stockholm, Sweden}
\ead{Jordan.Horowitz@umb.edu}

\begin{abstract}
In a thermodynamic process with measurement and feedback, the second law of thermodynamics is no longer valid.
In its place, various second-law-like inequalities have been advanced that each incorporate a distinct additional  term accounting for the information gathered through measurement.
We quantitatively compare a number of these information measures using an analytically tractable model for the feedback cooling of a Brownian particle. 
We find that the information measures form a hierarchy that reveals a web of interconnections.
To untangle their relationships, we address the origins of the information,  arguing that each information measure represents the minimum thermodynamic cost to acquire that information through a separate, distinct measurement protocol.
\end{abstract}



\submitto{New Journal of Physics}

\maketitle


\section{Introduction}

The Kelvin-Planck statement of the second law of thermodynamics forbids the existence of a cyclically operating device whose sole effect is to convert heat from a single thermal reservoir into an equal amount of work~\cite{Callen}.
However, we can circumvent this restriction, if our device operates via measurement and feedback: a possibility first envisioned by Szilard in his famous thought experiment~\cite{Leff}.
Recently, there has been renewed interest in this old idea  spurred by the development of a collection of distinct, second-law-like inequalities that quantify the interplay between the information gathered through measurement and the work that can be extracted in response through feedback.
For continuously operating devices at temperature $T$, all these predictions bound the extracted work rate ${\dot W}_{\rm ext}$ as
\begin{equation}\label{eq:2law}
{\dot W}_{\rm ext}\le k_{\rm B} T{\dot I},
\end{equation}
by some information acquisition rate, generically denoted here as ${\dot I}$, which differs in each second-law-like inequality, and $k_{\rm B}$ is Boltzmann's constant.
The first inequality of this form was derived by Sagawa and Ueda for a single feedback loop~\cite{Sagawa2008}, but subsequently has been extended  to include the repeated use of feedback, allowing for the application to continuously operating information engines ~\cite{Cao2009, Horowitz2010, Suzuki2010, Ponmurugan2010, Sagawa2011b, Abreu2012, Ito2013, Sandberg2014}.
In this case, the information rate is identified as the rate of growth of the transfer entropy~\cite{Schreiber2000} from the system to the measurement device (or feedback controller)~\cite{Sagawa2011b, Ito2013, Barato2013, Hartich2014}.
An alternative inequality identifies the information rate with the flow of mutual information between the system and a continuously-interacting auxiliary measurement device.
This information flow approach has been developed for small systems modeled as continuous diffusion processes~\cite{Allahverdyan2009}, discrete Markov jump processes~\cite{Hartich2014,Horowitz2014}, and for stochastic processes interacting discretly~\cite{Sagawa2012,Sagawa2013b}.
Yet another version has been suggested by Kim and Qian specifically for the feedback cooling of a harmonically-trapped Brownian particle, where the extracted work is bounded by a term they call entropy pumping~\cite{Kim2007}.
To date there is no clear information-theoretic interpretation of this term.
Nevertheless, this result conforms to the second-law-like structure in \eref{eq:2law}.
Further developments in this direction  are the inclusion of measurement errors and delay~\cite{Munakata2012,Munakata2013,Munakata2014}.
At first glance, this plethora of seeming similar predictions is confusing and raises questions about the interpretation as well as the utility of these information bounds.
To help clarify the situation, a number of studies have compared some of these measures from different points of view~\cite{Hartich2014,Allahverdyan2009,Horowitz2013,Barato2014}.
Our goal in this paper is to build on these works by providing a comprehensive, pedagogical comparison of all these information measures within a single framework in order show clearly their relationships and limitations.

There are essentially two ways to view \eref{eq:2law}.
The first is to treat \eref{eq:2law} simply as a numerical bound on the extracted work ${\dot W}_{\rm ext}$ without reference to the physical underpinnings of ${\dot I}$.
This is the point of view we typically take when investigating feedback (or information) engines~\cite{Suzuki2009,Abreu2011,Horowitz2011, Bauer2012,Horowitz2013,Horowitz2013b}, where our goal is to optimally extract the maximum amount of work; the maximum being any or all of the possible information measures.
In this respect, having so many bounds is problematic, since we are unsure which is the most appropriate.
Nevertheless, this is the approach we take in the first half our paper in \sref{sec:information}.
There we investigate the quantitative relationship between the various information measures by analytically calculating them in a Brownian particle model of feedback cooling, which we introduce in \sref{sec:model}.
We use this particular model, since it has been studied theoretically~\cite{Kim2007,Munakata2012,Munakata2013} and could be implemented experimentally in the setups of~\cite{Garnier2005, Joubaud2008}.
The analytical tractability of this model further lets us examine these information measures from the point of view of optimal control theory, which reveals intimate connections among them.
The second way to interpret \eref{eq:2law} is to take seriously its resemblance to the second law, and ask how far can we push this analogy?
In particular, the traditional statement of the second law dictates that the entropy production of the universe -- system {\it and} surroundings -- during a thermodynamic process must be positive~\cite{Callen}.
In feedback-driven systems, the surroundings not only include the traditional thermodynamic reservoirs, such as heat baths or chemical baths, but in addition they include an auxiliary system that records the measurement and feeds back that information.
In this case, does \eref{eq:2law} still represent the entropy production of the system and its surroundings, except now the surroundings contain the feedback device?
This is the question we address in the second half of our paper in \sref{sec:interpretation}.
There we observe that the transfer entropy rate and information flow have clear interpretations as the minimum entropy production required to acquire that information.
However, each one is associated with a different physical measurement scenario, that is  with a distinct surroundings in much the same way a particle reservoir differs from a thermal reservoir.


\section{Feedback cooling model}\label{sec:model}

Throughout, we will illustrate the different information concepts with a model for the feedback cooling of an underdamped Brownian particle~\cite{Kim2007,Munakata2012,Munakata2013}.
This will allow us to discuss each measure using the same language.
We therefore in this section introduce the dynamics of the model, both on the individual trajectory level and the ensemble level, as well as collect germane results regarding its energetics and thermodynamics.

\subsection{Dynamics, energetics, and thermodynamics without feedback}\label{sec:nofeedback}

Our quantity of interest is the time-dependent velocity $v_t$ of a trapped, underdamped Brownian particle of mass $m$, coupled to a thermal reservoir at temperature $T$ with viscous damping coefficient $\gamma$, evolving according to the Langevin equation~\cite{Kubo}
\begin{equation}\label{eq:Langevin0}
m{\dot v}_t=-\gamma v_t+f_t+\xi_t,
\end{equation}
where $f_t$ is an externally controlled force, and $\xi_t$ is zero-mean Gaussian white noise with covariance $\langle \xi_t\xi_s\rangle=2\gamma T\delta(t-s)$.
Starting here we  set Boltzmann's constant to  unity, $k_{\rm B}=1$.

In the absence of control, $f_t=0$, the velocity $v_t$ relaxes to an equilibrium Boltzmann distribution $p_{\rm eq}(v)\propto \exp[-mv^2/(2T)]$.
In the following, we will vary $f_t$ using feedback in order to cool the particle, that is  damp its thermal fluctuations, thereby reducing its kinetic temperature $T_{\rm kin}=m\langle v^2\rangle <T$.
Before we get to that, it is helpful to first review the energetics and thermodynamics of a driven, underdamped Brownian particle without feedback, so that we can appreciate the  differences that arise in the presence of feedback.

To this end, we require the Fokker-Planck equation associated with \eref{eq:Langevin0} for the time-dependent probability density $p_t(v)$~\cite{Risken},
\begin{eqnarray}\label{eq:FPv}
\partial_t p_t(v)& =-\partial_v J^v_t(v)\\
J^v_t(v)& =-\frac{1}{m}(\gamma v_t-f_t)p_t(v)-\frac{\gamma T}{m^2}\partial_vp_t(v).
\end{eqnarray}
where we have introduced the (probability) current $J^v_t$.
Anticipating our discussion of the thermodynamics, we divide the current into its irreversible half, which is anti-symmetric under time-reversal, and its reversible half, which is time-reversal symmetric, as~\cite{Risken,Spinney2012,Tome2010}
\begin{eqnarray}\label{eq:irrCurr}
J^{\rm irr}_{ t}(v)&=-\frac{\gamma}{m} vp_{ t}(v)-\frac{\gamma T}{m^2}\partial_vp_{ t}(v)\\
J^{\rm rev}_t(v)&=\frac{f_t}{m}p_t(v).
\end{eqnarray}
Key to this splitting is treating the force $f_t$ as even under time reversal, as typically assumed for a  force arising from an external potential.
With this identification, the irreversible portion of the current $J^{\rm irr}_t$ arises solely due to the forces imparted on the particle by its surroundings: the friction, $-\gamma v_t$, and the fluctuating force, $\xi_t$.

Moving on to the thermodynamics, we have from stochastic energetics an unambiguous identification of the heat flow into the system as the work done by the thermal reservoir on the particle~\cite{Munakata2012,Spinney2012,Sekimoto,Seifert2012}, which on average reads
\begin{equation}\label{eq:heat}
{\dot Q}=\int   mv J^{\rm irr}_{t}(v) \rmd v.
\end{equation}
It notably only depends on the irreversible current arising from the forces due to the thermal reservoir.
The particle's (internal) energy is its average kinetic energy
\begin{equation}
E=\left\langle \frac{1}{2}mv_t^2\right\rangle=\int \frac{1}{2}mv^2 p_t(v) \rmd v.
\end{equation}
By differentiating $E$ with time and substituting in the Fokker-Planck equation \eref{eq:FPv}, we are able  to identify the extracted work rate via the first law of thermodynamics ${\dot E}=-{\dot W}_{\rm ext}+{\dot Q}$,
\begin{equation}\label{eq:work2}
{\dot W}_{\rm ext}=-\langle f_tv_t\rangle,
\end{equation}
as the average power delivered against the external force $f_t$.

 From stochastic thermodynamics, we also have the (irreversible) entropy production rate~\cite{Spinney2012,Tome2010,Seifert2012}
\begin{equation}\label{eq:entNOFB}
{\dot S}_{\rm i}={\dot S}(v)-\frac{{\dot Q}}{T}=\frac{m^2}{\gamma T}\int \frac{[J^{\rm irr}_t(v)]^2}{p_t(v)} \rmd v\ge 0,
\end{equation}
where we have the traditional splitting into the time variation of the system's Shannon entropy $S(v)=-\int p_t(v)\ln p_t(v)\, \rmd v$,
\begin{equation}\label{eq:Shannon}
{\dot S}(v)=-\int J^v_t(v)\partial_v\ln p_t(v)\, \rmd v,
\end{equation}
 and the reversible entropy exchange with the environment
\begin{equation}
{\dot S}_{\rm env}=-\frac{{\dot Q}}{T}.
\end{equation}
Notably, the entropy production only depends on the irreversible current, since it is a measure of the time-reversal symmetry breaking of the dynamics~\cite{Spinney2012}.
This property is what allowed us to pullout the contribution due to the heat, which is also only a function of  the irreversible current.

\subsection{Dynamics and energetics with feedback}

Our main focus in this paper is feedback cooling, where we vary $f_t$ in response to measurements of the velocity.
 Following \cite{Munakata2013}, we consider a feedback protocol where we measure the velocity $v_t$ obtaining outcomes $y_t$ with some error, and then feed back those measurements by applying a force $f_t=-ay_t$ that acts as an additional friction, extracting work.
A simple way to incorporate measurement error is to add to our read-out of $v_t$ Gaussian white noise $\eta_t$ of zero mean and covariance $\langle \eta_t\eta_s\rangle=\sigma^2\delta(t-s)$, with $\sigma^2$ quantifying the measurement uncertainty: for example as $y_t=v_t+\eta_t$.
However, white noise fluctuations are very violent.
To make the problem more tractable, we smooth over the noise by applying a low-pass filter with time constant $\tau$ to the measurements:  $y_t=(1/\tau)\int_0^t  e^{-(t-s)/\tau}(v_s+\eta_s)\, \rmd s$~\cite{Astrom}.
We are therefore led to the following modified dynamics including measurement and feedback~\cite{Munakata2013}
\begin{equation}\label{eq:Langevin}
\eqalign{
m{\dot v}_t & =-\gamma v_t-ay_t+\xi_t\\
\tau {\dot y}_t & =-(y_t-v_t - \eta_t),}
\end{equation}
where $a$ is the feedback gain.
It is important to note at this point that $y_t$ is merely a model of measurement outcomes.
We are not making any assumption about the physical system that records the measurements, nor implements the feedback in response.

In general, the joint system relaxes to a time-independent, nonequilibrium steady state, where heat is continuously being extracted as work to maintain the particle at the cooled kinetic temperature.
This is the scenario we focus on in the following.

To discuss the energetics, we need the equivalent description of the dynamics in \eref{eq:Langevin}  in terms of the Fokker-Planck equation for the time-dependent probability density $p_t(v,y)$,
\begin{equation}\label{eq:FP}
\partial_tp_t(v,y)=-\partial_v J^v_t(v,y)-\partial_y J_t^y(v,y),
\end{equation}
with (probability) currents
\begin{equation}\label{eq:Curr}
\eqalign{
J_t^v(v,y)=-\frac{1}{m}(\gamma v+ay)p_t(v,y)-\frac{\gamma T}{m^2}\partial_vp_t(v,y)\\
J_t^y(v,y)=-\frac{1}{\tau}(y-v)p_t(v,y)-\frac{\sigma^2}{2\tau^2}\partial_yp_t(v,y).}
\end{equation}
Again we can split the velocity current $J^v_t$ into irreversible and reversible pieces, as in \eref{eq:irrCurr},
\begin{eqnarray}\label{eq:irrCurr2}
J^{\rm irr}_t(v,y)=-\frac{\gamma}{m}vp_t(v,y)-\frac{\gamma T}{m^2}\partial_vp_t(v,y) \\
J^{\rm rev}_t(v,y)=-\frac{a}{m}yp_t(v,y).
\end{eqnarray}
This splitting singles out the irreversible current as solely due to the thermal reservoir as before [cf.~\eref{eq:irrCurr}], which is required to correctly link the heat and entropy production in the following.
Again, this division relies on choosing $f_t=-ay_t$ as time-reversal symmetric, just as in the preceding section.

Our focus is the steady state solution, which due to the linear, Gaussian dynamics is the Gaussian probability density~\cite{Kubo},
\begin{equation}\label{eq:ssDist}
 p_{\rm s}(v,y)=\frac{1}{\sqrt{(2\pi)^2|{\bSigma}|}}\exp\left[{-\frac{1}{2}(v,y)\cdot{\bSigma}^{-1}\cdot(v,y)^T}\right],
\end{equation}
where the steady-state covariance matrix is
\begin{equation}\label{eq:Sigma}
\bSigma=
\left(\begin{array}{cc}
\sigma_{\rm v}^2 & \sigma_{\rm vy} \\
\sigma_{\rm vy} & \sigma^2_{\rm y}
\end{array}\right),
\end{equation}
and the associated steady-state currents are $J^v_{\rm s}$ and $J^y_{\rm s}$.
The entries of $\bSigma$ can be determined by plugging \eref{eq:ssDist} into \eref{eq:FP}, as detailed for a more general model in~\cite{Munakata2013}; however their precise expressions are unilluminating and therefore relegated to \ref{sec:ssDist}.
We do observe that the reduced distribution of the velocity $p_{\rm s}(v)=\int p_{\rm s}(v,y)\, \rmd y$ is also Gaussian.
Therefore, it has the same structure as an equilibrium distribution, but with a smaller variance, or a cooler effective temperature~\cite{Munakata2013}
\begin{equation}
T_{\rm kin}=T\frac{1+(a/\gamma)(a\sigma^2/(2T))+(1+a/\gamma)(\gamma\tau/m)}{1+a/\gamma+(1+a/\gamma)(\gamma\tau/m)} <T,
\end{equation}
where the inequality is only satisfied in the regime of good cooling, $a\sigma^2\le 2T$.
Otherwise too much measurement noise is fed back into the velocity, effectively heating it.

Again from stochastic energetics the heat current is identified as the energy lost due to the irreversible current arising from the thermal noise~\cite{Munakata2012,Sekimoto,Seifert2012}
\begin{equation}\label{eq:heat}
{\dot Q}=\int   mv J^{\rm irr}_{t}(v,y)\, \rmd v\rmd y=\int mv J^{\rm irr}_t(v)\, \rmd v,
\end{equation}
which importantly only depends on the velocity as in \eref{eq:irrCurr}, since the measurement and feedback do not affect the interaction with the thermal environment.
In a similar way as before \eref{eq:work2} the extracted work rate is
\begin{equation}
{\dot W}_{\rm ext}=a\langle y_tv_t\rangle,
\end{equation}
 due to the correlations between the feedback force and the particle.
In the steady state, ${\dot W}_{\rm ext}$ can be simplified using the defining equations for the elements of the covariance matrix $\bSigma$ in \ref{sec:ssDist},
\begin{equation}\label{eq:work}
{\dot W}_{\rm ext}=a\sigma_{\rm vy}=\frac{1}{\tau_{\rm v}}\left(T-T_{\rm kin}\right),
\end{equation}
in terms of the velocity's relaxation rate, $1/\tau_{\rm v}=\gamma/m$.
When the feedback is successful, and we have reduced the kinetic temperature $T_{\rm kin}<T$, we must be extracting work, ${\dot W}_{\rm ext}> 0$, recovering the results of~\cite{Munakata2013}.

We finally will require the fluctuating-trajectory solutions of \eref{eq:Langevin}  up to time $t$, $v_0^t=\{v_s\}_{s=0}^t$ and $y_0^t=\{y_s\}_{s=0}^t$.
We can obtain the probability densities for these trajectories by discretizing time and then using the usual procedure for obtaining path-integral densities, which we sketch  in \ref{sec:Disc}.
The joint density ${\mathcal P}[v_0^t,y_0^t]$ can be conveniently expressed in terms of two probability densities
\begin{equation}\label{eq:Phat1}
\hat{\mathcal P}[y_0^t|v_0^t,y_0]\propto\exp\left[-\int_0^t \rmd s\,\frac{(\tau{\dot y}_s+y_s-v_s)^2}{2\sigma^2}\right],
\end{equation}
suitably normalized,
and
\begin{equation}\label{eq:Phat2}
\hat{\mathcal P}[v_0^t|y_0^t,v_0]\propto\exp\left[-\int_0^t \rmd s\,\frac{(m{\dot v}_s+\gamma v_s+ay_s)^2}{4\gamma T/m^2}\right],
\end{equation}
as
\begin{equation}\label{eq:P}
{\mathcal P}[v_0^t,y_0^t]=\hat{\mathcal P}[v_0^t|y_0^t,v_0]\hat{\mathcal P}[y_0^t|v_0^t,y_0]p(v_0,y_0),
\end{equation}
with initial probability density $p(v_0,y_0)$.
It cannot be under emphasized that each $\hat{\mathcal P}$ is \emph{not} the conditional probability of the feedback process, \emph{i.e.}, $\hat{\mathcal P}[y_0^t|v_0^t,y_0]\neq {\mathcal P}[y_0^t|v_0^t,y_0]={\mathcal P}[v_0^t,y_0^t|y_0]/{\mathcal P}[v_0^t|y_0]$, since $v_t$ and $y_t$ influence each other when there is feedback~\cite{Sagawa2011b}.
Instead, we can understand $\hat{\mathcal P}[y_0^t|v_0^t,y_0]$ by first imagining that we fix the \emph{entire} velocity trajectory $v_0^t$, and then evolve $y_t$ alone according to \eref{eq:Langevin}.
This procedure has no feedback and the probability to observe a particular measurement trajectory is exactly $\hat{\mathcal P}[y_0^t|v_0^t,y_0]$.
A similar interpretation holds for $\hat{\mathcal P}[v_0^t|y_0^t,v_0]$ as well.
This distinction between ${\hat\mathcal P}$ and ${\mathcal P}$ will become important in \sref{sec:trans} when we introduce the transfer entropy rate.


\section{Information}\label{sec:information}

In this section, we present the definitions of the various measures of information that can be used to bound the extracted work during a feedback process.
In the next section, \sref{sec:interpretation}, we will discuss the physics behind them.

\subsection{Transfer entropy rate}\label{sec:trans}

The first information measure we discuss is the transfer entropy rate from $v_t$ to $y_t$.
The transfer entropy is a directional measure of information, which quantifies in an information-theoretic manner how much the dynamics (or more specifically the transition probabilities) of $y_t$ are influenced by $v_t$~\cite{Schreiber2000}.
For our continuous stochastic process, it reads
\begin{equation}\label{eq:trans}
{\dot I}_{\rm v\to y}=\lim_{t\to \infty}\frac{1}{t}\int\mathcal{D}[v_0^t]\mathcal{D}[y_0^t]\, {\mathcal P}[v_0^t,y_0^t]\ln\frac{\hat{\mathcal P}[y_0^t|v_0^t,y_0]}{{\mathcal P}[y_0^t|y_0]} \ge 0.
\end{equation}
In \ref{sec:Disc}, we justify this expression by discretizing the evolution and then utilizing the well-developed theory for repeated, discrete feedback~\cite{Sagawa2008,Horowitz2010, Ponmurugan2010,Hartich2014,Suzuki2009,Barato2013b}.
When no measurements are taking place, the dynamics of $y_t$ is independent of $v_t$, $\hat{\mathcal P}[y_0^t|v_0^t,y_0]={\mathcal P}[y_0^t|y_0]$,  and the transfer entropy rate is zero.
On the other hand, the more influence the velocity has on the measurement outcomes the larger the transfer entropy rate.
 Furthermore, when there is only one measurement the transfer entropy simplifies to the mutual information~\cite{Sagawa2011b}.
An alternative, equivalent expression for the transfer entropy rate in the context of continuous feedback has been introduced by Sandberg \emph{et al}~\cite{Sandberg2014}.
A similar analysis was performed by Fujitani and Suzuki for discrete Markov processes~\cite{Suzuki2010,Suzuki2009}.
The transfer entropy rate in feedback systems described by continuous-time, discrete Markov processes has been extensively studied in~\cite{Ito2013,Hartich2014,Barato2013b,Sagawa2011}.

To compare ${\dot I}_{\rm v\to y}$ with the other information measures, we calculate its value in our model of feedback cooling.
The calculation is facilitated by noting that for stationary Gaussian processes, as we have, integrals of the form \eref{eq:trans} can be conveniently expressed in terms of the power spectra -- Fourier transforms of the correlation functions.
For \eref{eq:trans}, we demonstrate in \ref{sec:power}  that it can be formulated as
\begin{equation}\label{eq:transOmega}
{\dot I}_{\rm v\to y}=-\frac{1}{4\pi}\int_{-\infty}^{\infty} \ln\frac{\hat{C}_{\rm yy|v}(\omega)} {{C}_{\rm yy}(\omega)}\,  \rmd \omega,
\end{equation}
where $C_{\rm yy}(\omega)$ is the power spectrum of $y_t$, and $\hat{C}_{\rm yy|v}(\omega)$ is the Fourier transform of the variance of $y_t$ given a fixed trajectory $v_0^t$.
We have carried out the integral in \ref{sec:Integral} with the result
\begin{equation}\label{eq:transCalc}
{\dot I}_{\rm v\to y}=\frac{\gamma}{2m} \left(\sqrt{1+\frac{2T/\gamma}{\sigma^2}}-1 \right)=\frac{1}{2\tau_{\rm v}}\left(\sqrt{1+\rm{SNR}}-1\right).
\end{equation}
New information is acquired at the relaxation rate of $v_t$, $\gamma/m=1/\tau_{\rm v}$; that is we learn new information about $v_t$ only as fast as $v_t$ changes enough to detect.
In addition, the transfer entropy rate does not depend on the feedback parameters $a$ and $\tau$, but only on the measurement accuracy $\sigma^2$ through  the dimensionless signal-to-noise ratio ${\rm SNR} = (2T/\gamma)/\sigma^2$, which quantifies the relative size of the measurement accuracy to the thermal diffusion of the velocity.
As a result, for perfect measurements without error, $\sigma=0$, the ${\rm SNR}$ diverges and with it the transfer entropy rate.
Thus, error-free measurement corresponds to infinite information, consistent with the notion that infinite information is required to localize a continuous variable with perfect precision.

\subsection{Information flow}\label{sec:bipartite}

We next consider the information flow, whose origin is in the exchange of information between the velocity and the auxiliary measurement device implementing the control.
It was first considered in the context of interacting diffusion processes~\cite{Allahverdyan2009}, but subsequently has been introduced in the analysis of the thermodynamics of continuously-coupled, discrete stochastic systems~\cite{Hartich2014,Horowitz2014,Shiraishi2014}.
When the coupling is not continuous, but each system takes turns evolving, the information flow simplifies to the mutual information \cite{Horowitz2014,Sagawa2012,Sagawa2013b}.
In order to facilitate connections to  the other information measures, we sketch in this section the basic arguments  leading to the information flow, following the program outlined in \cite{Horowitz2014}, and then calculate its value in our feedback cooling model.

First, we must note that strictly speaking this approach requires that $y_t$ be the degree of freedom of a physical system, not simply an abstract measurement outcome.
Still, in this section we would like not to comment on the precise thermodynamics of $y_t$, taking it only as a generic thermodynamic system.
We will come back to its precise interpretation in \sref{sec:interpretation} when we compare the physics underlying the different information measures.

The key insight in this approach is that the (irreversible) entropy production of the joint system of $v_t$ and $y_t$ can be divided as
\begin{equation}\label{eq:Ssplit}
{\dot S}_{\rm i}={\dot S}_{\rm i}^v+{\dot S}_{\rm i}^y\ge 0,
\end{equation}
with positive contributions arising due to the irreversible current in the $v$-direction~\eref{eq:irrCurr2},
\begin{equation}\label{eq:entV}
{\dot S}^v_{\rm i}=\frac{m^2}{\gamma T}\int \frac{[J^{\rm irr}_{t}(v,y)]^2}{p_{t}(v,y)}\, \rmd v\rmd y\ge 0,
\end{equation}
and  separately from $y_t$, ${\dot S}_{\rm i}^y$.
The next step is to perform the traditional splitting of ${\dot S}^v_{\rm i}$ into the variation of the Shannon entropy due to $v_t$ [cf.~\eref{eq:Shannon}],
\begin{equation}\label{eq:sysEntV}
{\dot S}(v)=-\int J^v_t(v,y)\partial_v \ln p_t(v)\, \rmd v\rmd y,
\end{equation}
and the heat ${\dot Q}$ \eref{eq:heat} as
\begin{equation}\label{eq:2lawBi}
{\dot S}^v_{\rm i}={\dot S}(v)-\frac{\dot Q}{T}+{\dot I}_{\rm flow}\ge 0.
\end{equation}
The additional contribution due to the influence of $y_t$ is an information-theoretic piece
\begin{equation}
{\dot I}_{\rm flow}=-\int J_{t}^v(v,y)\partial_v\ln \frac{p_{ t}(v,y)}{p_{ t}(v)p_{ t}(y)}\, \rmd v \rmd y,
\end{equation}
which is (minus) the variation of the mutual information~\footnote{We have defined the information flow with the opposite sign convention of \cite{Hartich2014,Allahverdyan2009,Horowitz2014,Shiraishi2014}, so that it is positive in the cooling regime, allowing a straightforward comparison to the other information measures.}
\begin{equation}
I(v_t;y_t)=\int p_t(v,y)\ln \frac{p_t(v,y)}{p_t(v)p_t(y)}\,\rmd v\rmd y
\end{equation}
between $v_t$ and $y_t$,  due to the fluctuations of $v_t$~\cite{Cover}.
The mutual information $I(v_t;y_t)$ is a measure of correlations, quantifying how knowledge of the measurement outcomes reduces uncertainty in the velocity.
While ${\dot I}_{\rm flow}$ may be positive or negative, in the regime of good cooling where we are extracting work, we will always have ${\dot I}_{\rm flow}\ge0$.
In the steady state, $J^v_{\rm s}=0$, and ${\dot Q}={\dot W}_{\rm ext}$, so that \eref{eq:2lawBi} reduces to~\cite{Hartich2014,Horowitz2014}
\begin{equation}
{\dot W}_{\rm ext}\le T{\dot I}_{\rm flow},
\end{equation}
in the form of \eref{eq:2law}.

Employing the steady-state solution in \eref{eq:ssDist}, we have for the steady-state information flow
\begin{equation}
\label{eq:infoflow}
 {\dot I}_{\rm flow}=\frac{\gamma}{m}\left(\frac{T}{m}\frac{\sigma_{\rm y}^2}{|\bSigma|}-1\right)\ge 0.
\end{equation}
where $|\bSigma|$ denotes the determinant of $\bSigma$.
Unfortunately, we have been unable to formulate a more transparent expression in general.
 Even still, the information rate again only grows as fast as the relaxation rate of the velocity $\gamma/m=1/\tau_{\rm v}$.

\subsection{Entropy pumping}

For the feedback cooling of a Brownian particle without errors an entropy pumping bound has been  introduced by Kim and Qian~\cite{Kim2007}.
This approach has subsequently been developed by Ge~\cite{Ge2014} and extended to the setup in \eref{eq:Langevin} by Munakata and Rosinberg~\cite{Munakata2012,Munakata2013,Munakata2014}, which we discuss in this section.

The entropy pumping approach is based on a coarse graining of the Fokker-Planck equation \eref{eq:FP}.
Following \cite{Munakata2013}, we formally integrate out $y_t$ from \eref{eq:FP} to obtain the reduced Fokker-Planck equation
\begin{equation}\label{eq:FPcg}
\partial_tp_t(v)=\partial_v\left(\frac{1}{m}\left(\gamma v+{\tilde f}^{\rm fb}_t(v)\right)p_t(v)+\frac{\gamma T}{m^2}\partial_vp_t(v)\right)=-\partial_v {\tilde J}_t(v),
\end{equation}
where we have identified an effective feedback force
\begin{equation}
\tilde{f}^{\rm fb}_t(v)=a\int  y p_t(y|v)\rmd y.
\end{equation}
Furthermore, we treat ${\tilde f}^{\rm fb}_t$ as time-reversal symmetric, as we would expect for an external force~\cite{Munakata2013}.
 In which case, we single out from the coarse-grained current the irreversible current exactly as for the no-feedback case~\eref{eq:irrCurr},
\begin{equation}\label{eq:coarseCurr}
{\tilde J}_t(v)=J^{\rm irr}_t(v)-\frac{1}{m} {\tilde f}^{\rm fb}_t(v).
\end{equation}
This will allow us to connect the entropy production in the environment with the heat.

\Eref{eq:FPcg} is not a closed equation for $p_t(v)$; the measurement dynamics are required to solve it.
Nevertheless, the entropy pumping approach is to treat \eref{eq:FPcg} as a thermodynamically consistent equation for $p_t(v)$ with an effective external force ${\tilde f}^{\rm fb}_t$.
In this case, the entropy balance is developed in analogy to the no-feedback setup, as in \eref{eq:entNOFB},
\begin{equation}\label{eq:entVcg}
\dot{\tilde S}^v_{\rm i}=\frac{m^2}{\gamma T}\int  \frac{[J^{\rm irr}_{t}(v)]^2}{p_{t}(v)}\rmd v=\dot{ S}(v)-\frac{\dot Q}{T}+{\dot I}_{\rm pump}\ge 0,
\end{equation}
where the second equality follows by substituting in definition of the coarse-grained current ${\tilde J}_t(v)$ in \eref{eq:coarseCurr}.
Here, ${\dot S}(v)=-\int {\tilde J}_t(v)\partial_v\ln p_t(v)\, \rmd v$ is equivalent to the expression for the rate of  change of the system's Shannon entropy including feedback in \eref{eq:sysEntV}, and the additional entropy pumping term arises due to the coarse-grained feedback force,
\begin{equation}
{\dot I}_{\rm pump}=\int  p_t(v)\partial_v\frac{1}{m}{\tilde f}^{\rm fb}_{t}(v)\rmd v.
\end{equation}
As pointed out in \cite{Munakata2013}, the feedback force is proportional to the minimum mean square error estimate of $y_t$ given $v_t$.
Other than that though, there does not appear to be a crisp  interpretation of the entropy pumping as a form of information, like for the transfer entropy rate and information flow.

Using  the steady-state distribution in \eref{eq:ssDist}, we have for the steady-state entropy pumping~\cite{Munakata2013}
\begin{eqnarray}
 {\dot I}_{\rm pump}=\int  p_{\rm s}(v)\partial_v\frac{1}{m}{\tilde f}^{\rm fb}_{\rm s}(v)\rmd v&=\frac{a}{m}\frac{\sigma_{\rm vy}}{\sigma_{\rm v}^2}=\frac{1}{\tau_{\rm v}}\left(\frac{T-T_{\rm kin}}{T_{\rm kin}}\right)\ge 0,
\end{eqnarray}
with positivity guaranteed when there is cooling $T\ge T_{\rm kin}$.

\subsection{Trajectory mutual information}\label{sec:traj}

Another information measure that has aroused some attention  is the mutual information rate between the entire $v_0^t$ and $y_0^t$ trajectories~\cite{Barato2013,Diana2013b}.
For continuous stochastic processes, the trajectory mutual information rate is~\cite{Cover}
\begin{equation}\label{eq:traj}
\fl{\dot I}_{\rm traj}=\lim_{t\to \infty}\frac{1}{t}I(v_0^t;y_0^t)=\lim_{t\to \infty}\frac{1}{t}\int\mathcal{D}[v_0^t]\mathcal{D}[y_0^t]\, {\mathcal P}[v_0^t,y_0^t]\ln\frac{{\mathcal P}[v_0^t,y_0^t]}{{\mathcal P}[v_0^t]{\mathcal P}[y_0^t]}\ge 0.
\end{equation}
It quantifies how much the uncertainty about the entire velocity trajectory $v_0^t$ is reduced given knowledge of the entire measurement trajectory $y_0^t$, and vice versa, as it is symmetric.

The ${\dot I}_{\rm traj}$ bound on the extracted work follows readily once we observe a close connection between the trajectory mutual information and the transfer entropy pointed out in \cite{Hartich2014};
by substituting  ${\mathcal P}$ with $\hat{\mathcal P}$ \eref{eq:P} in ${\dot I}_{\rm traj}$, it follows that
\begin{equation}\label{eq:trajTrans}
{\dot I}_{\rm traj}={\dot I}_{\rm v\to y}+{\dot I}_{\rm y\to v},
\end{equation}
after identifying the transfer entropy rate from $y_t$ to $v_t$, ${\dot I}_{\rm y\to v}\ge 0$, defined analogously to ${\dot I}_{\rm v\to y}$ \eref{eq:trans}.
The positivity of the transfer entropy implies that
\begin{equation}\label{eq:2lawTraj}
{\dot W}_{\rm ext}\le T{\dot I}_{\rm v\to y}\le T{\dot I}_{\rm traj},
\end{equation}
giving \eref{eq:2law} for the trajectory information, which is always weaker than the transfer entropy bound.

The trajectory information rate has been studied in numerous contexts and has a well-known expression in terms of power spectra~\cite{Pinsker,Fano,Munakata2006,Tostevin2010} that we recall in~\ref{sec:power},
\begin{equation}\label{eq:trajOmega}
{\dot I}_{\rm traj}=-\frac{1}{4\pi}\int_{-\infty}^{\infty} \ln\left(1-\frac{|C_{\rm vy}(\omega)|^2}{C_{\rm vv}(\omega)C_{\rm yy}(\omega)}\right) \rmd \omega.
\end{equation}
In \ref{sec:Integral}, we perform this integral to find
\begin{equation}\label{eq:traj2}
\eqalign{
{\dot I}_{\rm traj}&=\frac{\gamma}{2m}\left(\sqrt{1+\frac{2T/\gamma}{\sigma^2}}-1\right) + \frac{1}{2\tau}\left(\sqrt{1+\frac{a^2\sigma^2}{2\gamma T}}-1\right).
}
\end{equation}
Comparing with \eref{eq:trajTrans}, we have as a byproduct the transfer entropy rate from $y_t$ to $v_t$,
\begin{equation}
{\dot I}_{\rm y\to v}=\frac{1}{2\tau}\left(\sqrt{1+\frac{a^2\sigma^2}{2\gamma T}}-1\right).
\end{equation}

\subsection{Maximum work}

A final  bound on the extracted work  is simply to maximize ${\dot W}_{\rm ext}$ in  \eref{eq:work} with respect to the measurement parameters $a$ and $\tau$.
While the result is not general, remarkably for  linear Guassian processes it has a close connection
with the transfer entropy rate, as first noticed by Sandberg \emph{et al}~\cite{Sandberg2014}.
Using standard calculus methods, the extracted work is bounded above by its maximal value
\begin{equation}\label{eq:maxWork}
{\dot W}_{\rm ext}\le {\dot W}^{\rm max}_{\rm ext}=T_{\rm kin}{\dot I}_{\rm v\to y},
\end{equation}
akin to \eref{eq:2law}, for parameter values
\begin{equation}\label{eq:maxWorkparameters}
a^*=\gamma\left(\sqrt{1+\frac{2 T/\gamma}{ \sigma^2}}-1\right), \qquad \tau^*=0.
\end{equation}
The optimal measurement has no low-pass filtering: It is immediately fed back into the particle to control it.
Remarkably, the extracted work is again bounded by the transfer entropy rate, except multiplied by the cooled kinetic temperature of the particle, instead of $T$.

\subsection{Discussion}

\subsubsection{Quantitative comparison of information measures.}
\begin{figure}[htb]
\centering
\includegraphics[scale=.43]{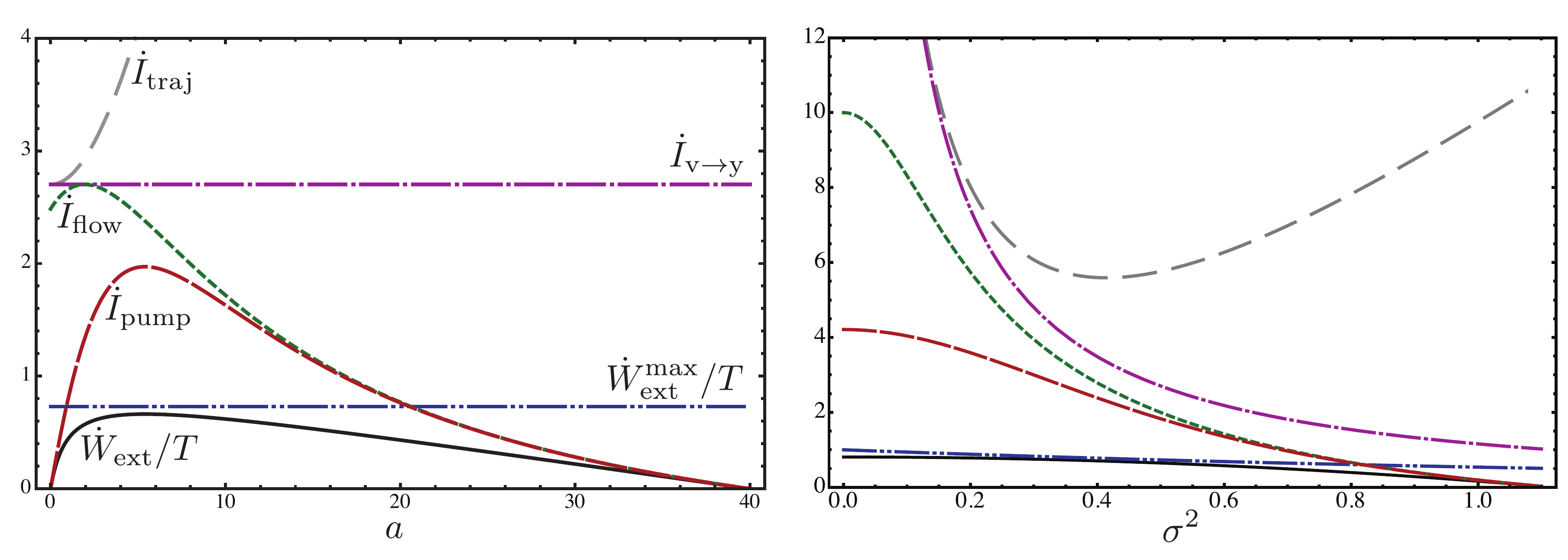}
\caption{Comparison of the information measures and the extracted work ${\dot W}_{\rm ext}/T$ in dimensionless units as a function of the feedback gain $a$ (left) and measurement error $\sigma^2$ (right). Parameters $m=1$, $\gamma=1$, $T=5$, $\tau=0.1$,  with  $\sigma=0.5$ (left) and $a=8$ (right)}.
\label{fig:info}
\end{figure}

To better understand the relationship between all of these information measures, we plot them all together with ${\dot W}_{\rm ext}$ in \fref{fig:info} as a function of the feedback gain $a$ and measurement error $\sigma^2$ in the range where cooling occurs (${\dot W}_{\rm ext}\ge0$).
As expected, each information measure bounds the extracted work.

The most striking feature of \fref{fig:info} is the hierarchy of information measures,
\begin{equation}\label{eq:Iineq}
 {\dot I}_{\rm traj}\ge  {\dot I}_{\rm v\to y}\ge {\dot I}_{\rm flow}\ge {\dot I}_{\rm pump},
\end{equation}
apart from ${\dot W}_{\rm ext}^{\rm max}$, which does not actually have a generic information interpretation.
In fact, this ranking  holds quite generally.
We have already seen that ${\dot I}_{\rm traj}\ge {\dot I}_{\rm v\to y}$ in \sref{sec:traj} when discussing the second-law-like inequality for the trajectory information.
 The middle inequality, ${\dot I}_{\rm v\to y}\ge {\dot I}_{\rm flow}$,  has been demonstrated by Hartich \emph{et al}~\cite{Hartich2014} for continuous-time, discrete Markov jump processes.  
For diffusion processes, a similar conclusion was reached by Allahverdyan \emph{et al}~\cite{Allahverdyan2009} except for a slightly different transfer entropy rate that uses only the most recent measurement, which upper bounds the transfer entropy rate considered here, as pointed out in~\cite{Hartich2014}.
Nevertheless, the proof for jump processes in~\cite{Hartich2014} can be carried over to diffusion processes, once their evolution is discretized.
The last inequality between the information flow and the entropy pumping also is generic.
This follows by bounding the steady-state entropy production of $v_t$ in the information-flow description \eref{eq:entV} using a coarse-graining inequality~\cite{Munakata2013} to connect it to the coarse-grained, entropy-pumping approach:
\begin{eqnarray}
\fl {\dot S}^v_{\rm i}=-\frac{{\dot Q}}{T}+{\dot I}_{\rm flow}=\frac{m^2}{\gamma T}\int \frac{[J^{\rm irr}_{\rm s}(v,y)]^2}{p_{\rm s}(v,y)}\rmd v\rmd y\ge \frac{m^2}{\gamma T}\int  \frac{[J^{\rm irr}_{\rm s}(v)]^2}{p_{\rm s}(v)}\rmd v=-\frac{\dot Q}{T}+{\dot I}_{\rm pump}=\dot{\tilde S}_{\rm i}^v,
\end{eqnarray}
where we have employed the entropy balance of entropy pumping in \eref{eq:entVcg}.
Clearly,
\begin{equation}\label{eq:biPuBounds}
{\dot I}_{\rm flow}\ge {\dot I}_{\rm pump}.
\end{equation}
As a lower bound on all other information measures, the entropy pumping can be given an information-theoretic interpretation, which till now has been lacking, as a minimal information requirement for successful feedback cooling.

An alternative perspective on this hierarchy of information measures comes from considering the efficiency of work extraction
\begin{equation}\label{eq:eff}
\varepsilon=\frac{{\dot W}_{\rm ext}}{T{\dot I}}\le 1.
\end{equation}
By utilizing the smaller information measures, we will estimate higher efficiencies, even without changing the measurement or feedback procedure.
This conclusion is somewhat surprising, since it makes the notion of efficiency somewhat arbitrary.
We will come back to this observation later, after discussing the physical origins of the different information measures.

 We also see in figure~\ref{fig:info} that the transfer entropy rate and the trajectory mutual information diverge as the measurement error tends to zero, $\sigma^2\to 0$; whereas the other measures remain finite.
Munakata and Rosinberg have also observed that the entropy pumping displays a nontrivial structure, attaining a maximum at  the maximum cooling rate~\cite{Munakata2013}.
Figure~\ref{fig:info} demonstrates that ${\dot I}_{\rm flow}$ displays a similar structure, but its maximum does not quite correspond to the maximum cooling.
Most likely, this discrepancy arises due to the effect of coarse-graining.

\subsubsection{Optimal control and the Kalman-Bucy filter.}\label{sec:kalman}
Interestingly, closer connections exist between the information flow, transfer entropy rate, and maximum extractable work that are  revealed by re-examining our feedback problem from the perspective of optimal control theory.

The  feedback cooling we have been addressing is a special case of a classic problem in optimal control theory: the characterization of feedback controllers that minimize quadratic performance objectives of the form
\begin{equation}
{\mathcal J}=\langle v_t^2 \rangle + \rho \langle f_t^2  \rangle,
\end{equation}
where $\rho > 0$ is a constant parameter used to tune the trade-off between keeping small fluctuations in $v_t$ and applying a strong control force $f_t$, for example~\cite{Astrom2}.
For the special case of cooling, we have been focused on minimizing $\langle v^2_t\rangle$ alone, which corresponds to $\rho \to 0$.

Assuming linear dynamics and Gaussian noise, the optimal feedback controller with access to noisy measurements $v_t+\eta_t$ can be written in the form
\begin{equation} \label{eq:optcontrol}
\eqalign{
m\dot{\hat v}_t&=-\gamma\hat{v}_t-G\hat{v}_t+K(v_t+\eta_t-{\hat v}_t)\\
f_t &= - G\hat v_t,
}
\end{equation}
where $\hat v_t$ is the abstract dynamical state of the controller, and $G$ and $K$ are carefully chosen constants.
According to the separation principle~\cite{Astrom2,wonham68}, these parameters $G$ and $K$ can be determined as the solutions to two independent optimization problems: the optimal gain $G$ is obtained by minimizing ${\mathcal J}$, temporarily assuming there is no measurement noise, $\sigma=0$; whereas the optimal $K$ is obtained by minimizing the estimation error, see below, and is independent of the tuning parameter $\rho$.
While the exact expression for the optimal gain $G$ is of little interest to us here, we do note that it tends monotonically to infinity as $\rho \rightarrow 0$.
This makes intuitive sense, since $\rho\rightarrow 0$ means we only care about minimizing the variance $\langle v_t^2\rangle$ and assess no cost for large control forces $\langle f_t^2\rangle$.
On the other hand, optimal filtering theory selects an optimal $K$  by minimizing the estimation error,
\begin{equation}\label{eq:optest}
{\mathcal E}_t \equiv \min_K \langle (v_t - \hat v_t)^2\rangle,
\end{equation}
 given all the past measurements $(v+\eta)_0^t$. The steady-state optimum, achieved for
 \begin{equation}
 \label{eq:optimalK}
K = \gamma\left(\sqrt{1+\frac{2 T/\gamma}{ \sigma^2}}-1\right),
\end{equation}
is
\begin{equation}
{\mathcal E} = \frac{\sigma^2}{m}K = \sigma^2\frac{\gamma}{m}\left(\sqrt{1+\frac{2 T/\gamma}{ \sigma^2}}-1\right).
\end{equation}
 Thus, $\hat v_t$ represents the best estimate of  $v_t$ given all past measurements. In fact, no other filter, nonlinear or otherwise, can produce a better estimate than the one described here, which is known as the Kalman-Bucy filter \cite{Astrom2,Bucy}.

Remarkably, the optimal controller~\eref{eq:optcontrol} with Kalman-Bucy filter can always be realized using the feedback cooling dynamics in \eref{eq:Langevin} by a simple rescaling
\begin{equation}
{\hat v}_t=\left(\frac{K}{\gamma+K+G}\right)y_t,
\end{equation}
and choosing the parameters $a$ and $\tau$ as
\begin{equation}\label{eq:KBparameters}
a_{\rm KB}=\frac{G K}{\gamma+K+G},\qquad \tau_{\rm KB}=\frac{m}{\gamma+K+G}.
\label{eq:optrealized}
\end{equation}
This mapping allows us to investigate our information measures from a new point of view by replacing $y_t$ with the optimal ${\hat v}_t$.

For starters, maximal cooling, which coincides with the maximum extracted work ${\dot W}_{\rm ext}^{\rm max}$~\eref{eq:maxWork}, is obtained when $G\to\infty$, in which case the optimal controller~\eref{eq:KBparameters} becomes
\begin{equation}
a_{\rm KB}= K= \gamma\left(\sqrt{1+\frac{2 T/\gamma}{ \sigma^2}}-1\right), \qquad \tau_{\rm KB}=0,
\end{equation}
recovering $a^*$ and $\tau^*$ in \eref{eq:maxWorkparameters} as expected.

The optimal controller also extracts the maximum amount of information.
To see this, first note that optimality of the estimate ${\hat v}_t$ implies that the estimation error is stochastically orthogonal to the estimate $\langle {\hat v_t}(v_t-{\hat v}_t)\rangle=0$ for all $t$~\cite{Astrom2}.
This property greatly simplifies the steady-state covariance matrix
\begin{equation}\label{eq:optCov}
\bSigma=
\left(\begin{array}{cc}
\sigma_{\rm v}^2 & \sigma_{\rm v {\hat v}} \\
\sigma_{\rm v{\hat v}} & \sigma^2_{\rm {\hat v}}
\end{array}\right) =
\left(\begin{array}{cc}
\sigma^2_{\rm {\hat v}}+{\mathcal E} &    \sigma^2_{\rm {\hat v}}  \\
\sigma^2_{\rm {\hat v}} & \sigma^2_{\rm {\hat v}}
\end{array}\right),
\end{equation}
where the variance of the estimate is simply
\begin{equation}
\sigma^2_{\rm {\hat v}}= \frac{K^2 \sigma^2}{2m(\gamma + G)}.
\end{equation}
Note optimal cooling is achieved by $G\rightarrow \infty$, forcing  $\sigma_{\rm{\hat v}}^2\to 0$, so that fluctuations in the velocity  $\sigma_{\rm v}^2={\mathcal E}$ are only caused by estimation error.
Furthermore, by exploiting the structure of $\bSigma$ in \eref{eq:optCov}, the expression for the
steady-state information flow  \eref{eq:infoflow} greatly simplifies,
\begin{equation}\label{eq:Iequals}
 \fl{\dot I}_{\rm flow} =\frac{\gamma}{m}\left(\frac{T}{m}\frac{\sigma_{\rm {\hat v}}^2}{|\bSigma|}-1\right)
 = \frac{\gamma}{m}\left(\frac{T}{m{\mathcal E}}-1\right) = \frac{\gamma}{2m}\left(\sqrt{1+\frac{2T/\gamma}{ \sigma^2}}-1\right) = {\dot I}_{\rm v\to y},
\end{equation}
for all $G$.
This is a very interesting observation, supporting the claimed optimality of the Kalman-Bucy filter.
We already know that ${\dot I}_{\rm flow} \leq {\dot I}_{\rm v\to y}$.
What we see here is that the class of
controllers given by (\ref{eq:optcontrol}), \emph{i.e.}, with $K$ fixed (\ref{eq:optimalK}) and $G$ free, saturates the bound,  maximizing the information flow.
Hence, a controller with a small gain $G$ (zero even) only uses  information to create an optimal estimate of the process, whereas a high gain cools as well.
To gain further insight into equality \eref{eq:Iequals}, we have to look at the transfer entropy rate and information flow from a different perspective.
Namely, the transfer entropy rate can also be defined as the rate of growth of the mutual information between $v_t$ and the entire trajectory of measurement outcomes $y_0^t$, that is the change in $I(v_t;y_0^t)$.
On the other hand, the information flow is the rate of growth of  the mutual information between $v_t$ and just the most recant measurement $y_t$, that is the change in $I(v_t;y_t)$.
The inequality ${\dot I}_{\rm flow} \le {\dot I}_{\rm trans}$ is then related to the simple idea that the entire trajectory of measurements contains more information than just the last.
Now, it is known that the Kalman-Bucy filter ${\hat v}_t$ is a sufficient statistic for the conditional distribution of $v_t$ given the measurements $y_0^t$~\cite{Mitter+05}.
In other words, everything useful in a collection of measurements for predicting $v_t$ is contained in just ${\hat v}_t$, or in terms of the mutual information $I(v_t;{\hat v}_0^t)=I(v_t;{\hat v}_t)$.
This equality translated into rates implies \eref{eq:Iequals}.

\begin{figure}[htb]
\centering
\includegraphics[scale=.5]{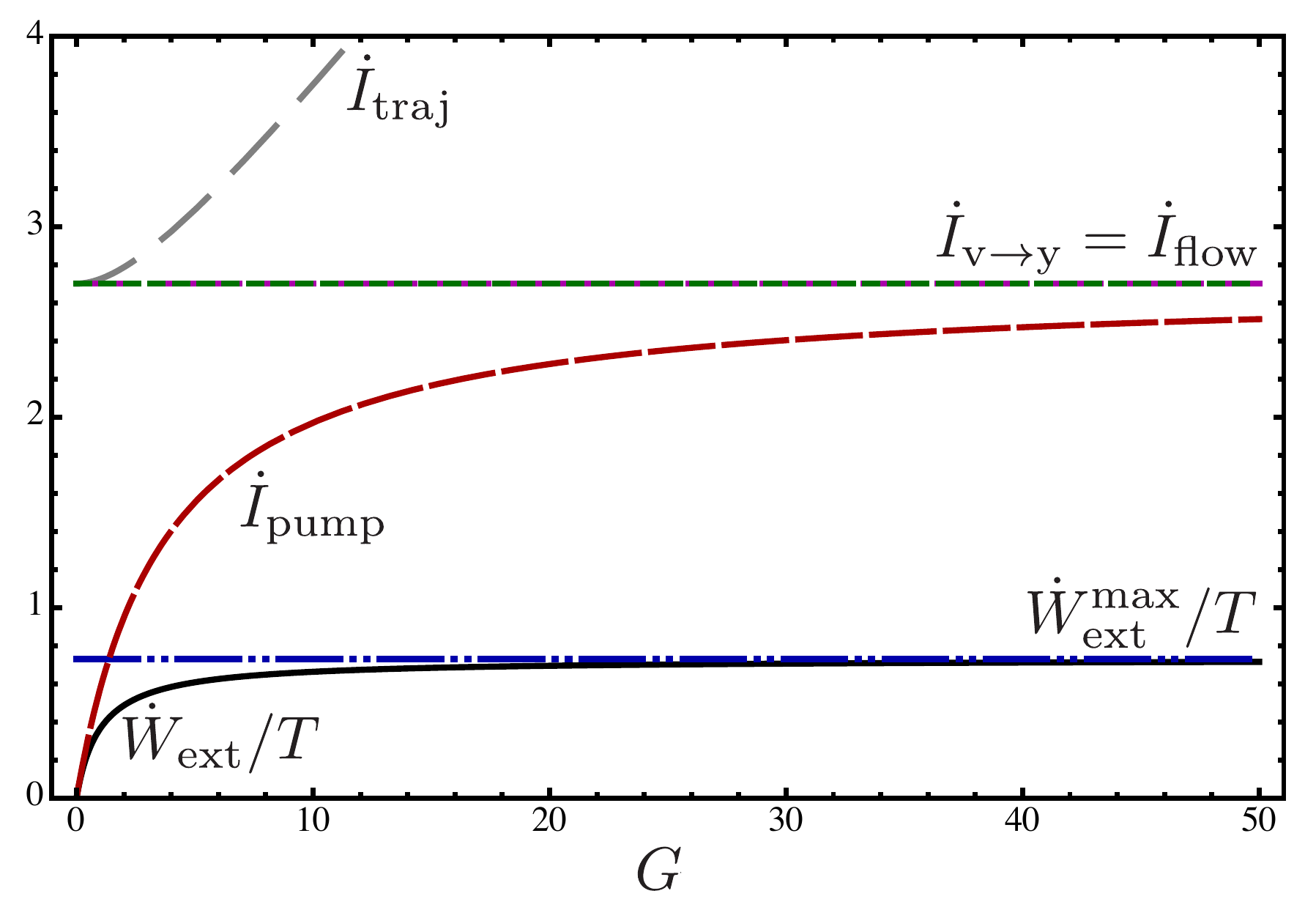}
\caption{Comparison of the information measures for the Kalman-Bucy filter with the extracted work ${\dot W}_{\rm ext}/T$ in dimensionless units as a function of the Kalman-Bucy gain $G$ with parameters $m=1$, $\gamma=1$, $T=5$, $\sigma=0.5$, and $\tau=0.1$.}
\label{fig:kalman}
\end{figure}

In figure~\ref{fig:kalman}, we illustrate how the extracted work depends on $G$, and how the maximum is asymptotically achieved.
 In addition, we see that ${\dot I}_{\rm flow} = {\dot I}_{\rm v\to y}$ holds for all $G$.
We can also conclude that with certain choices of $a$ and $\tau$ (namely $a_{\rm KB}$ and $\tau_{\rm KB}$ in \eref{eq:KBparameters})  our original setup \eref{eq:Langevin} can always saturate ${\dot I}_{\rm flow} \leq {\dot I}_{\rm v\to y}$, which is indeed observed in figure~\ref{fig:info} for $a\approx 2$.


\section{Energetics of Information and Measurement}\label{sec:interpretation}

We have seen that there are various, distinct measures of information that each offer a nontrivial bound for the extracted work.
However, there does not seem to be an obvious reason to prefer any of one these measures.
To this end, we investigate their origins in this section.
We will find that the transfer entropy rate and the information flow  both correspond to the information that is recorded in an auxiliary system, or memory, and therefore is subject to the limits of thermodynamics, as originally suggested by Landauer~\cite{Leff}.
In particular, we show that these two information measures both bound the minimum energy required to gather that information through distinct thermodynamic processes, implying that the energy that we are able to extract as work originates in the (free) energy supplied by the memory.

\subsection{Information flow}\label{sec:bipartiteMeas}

\begin{figure}
\centering
\includegraphics[scale=.6]{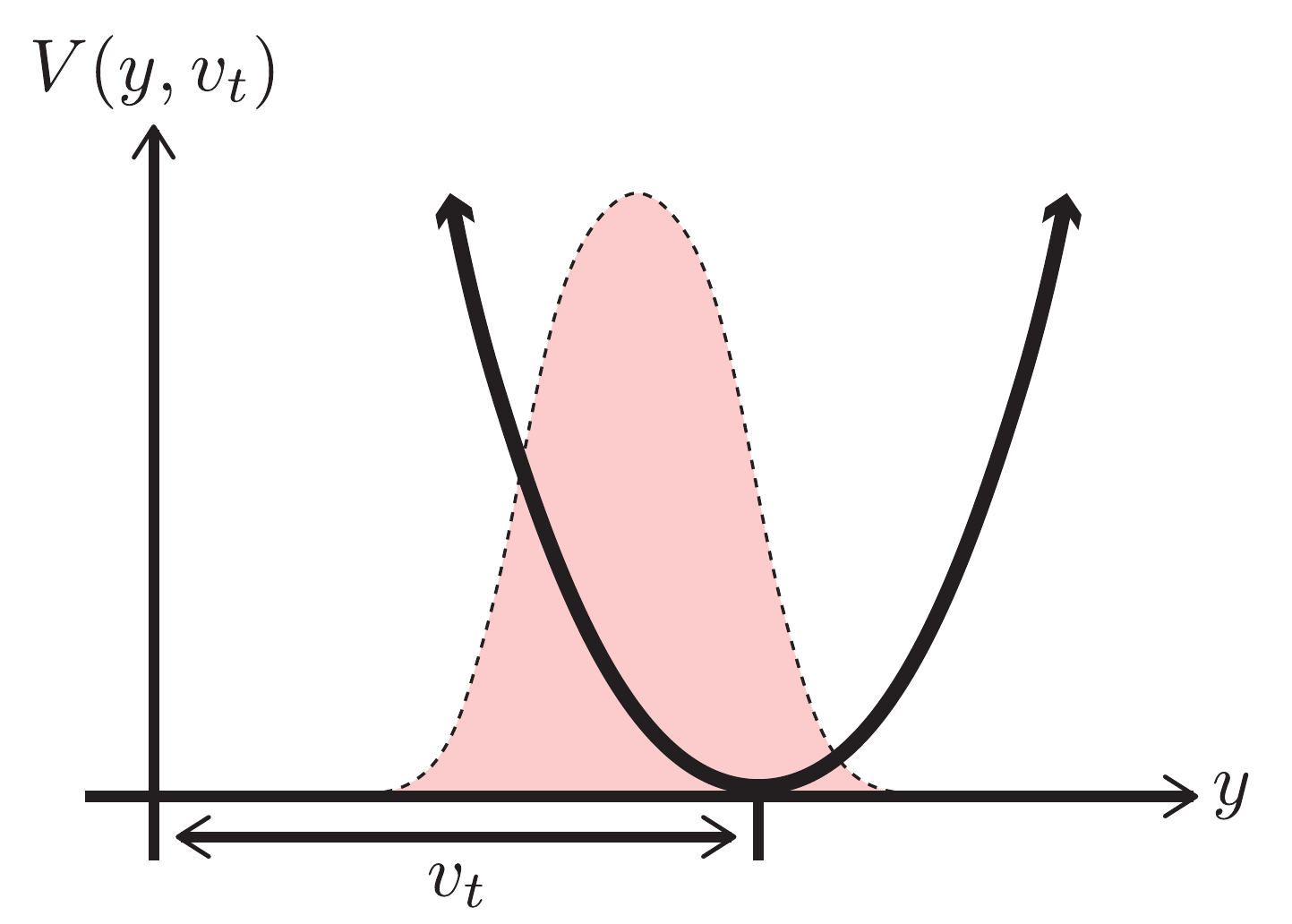}
\caption{Illustration of the moving potential experienced by the memory degree of freedom, $V(y,v_t)=(y-v_t)^2/2$, centered about the time-dependent velocity $v_t$.  The instantaneous probability density of $y_t$ (shaded pink region) is Gaussian and lags behind the potential due to the finite relaxation time $\tau$. }
\label{fig:flowPotential}
\end{figure}

Let us start with the simpler measurement scenario corresponding to the information flow ${\dot I}_{\rm flow}$.
Actually, we have already touched on its physical interpretation when we introduced it in \sref{sec:bipartite}.
Recall, there we considered the measurement outcomes $y_t$ to correspond to a physical degree of freedom of an auxiliary system.
We now clarify that interpretation by taking $y_t$ to be the position of a secondary, harmonically-trapped, \emph{overdamped} Brownian particle.
To be thermodynamically consistent, the origin of the measurement noise must be a thermal reservoir, which requires imposing the Fluctuation-Dissipation theorem~\cite{Seifert2012}:
\begin{equation}
\sigma^2=2\tau T.
\end{equation}
We have chosen the temperature of the measurement device to be the same as the controlled system, which is the customary choice.
From this point of view, \eref{eq:Langevin} is the equation of motion for an overdamped Brownian particle with viscous damping coefficient $\tau$ trapped in a harmonic potential $V(y,v)=(y-v)^2/2$ of unit spring constant,  centered about the velocity, as illustrated in \fref{fig:flowPotential}.
 Alternatively, such a coupling can be implemented in an electric circuit as was presented in \cite{Sandberg2014}. The result is that the  position of the measurement oscillator $y_t$ feels a fluctuating force making it track the velocity $v_t$, thereby establishing and maintaining correlations.
Roughly speaking, the measurement oscillator is constantly learning new information at a rate ${\dot I}_{\rm flow}$, which keeps getting rewritten in the value of its position.

\begin{figure}[htb]
\centering
\includegraphics[scale=.5]{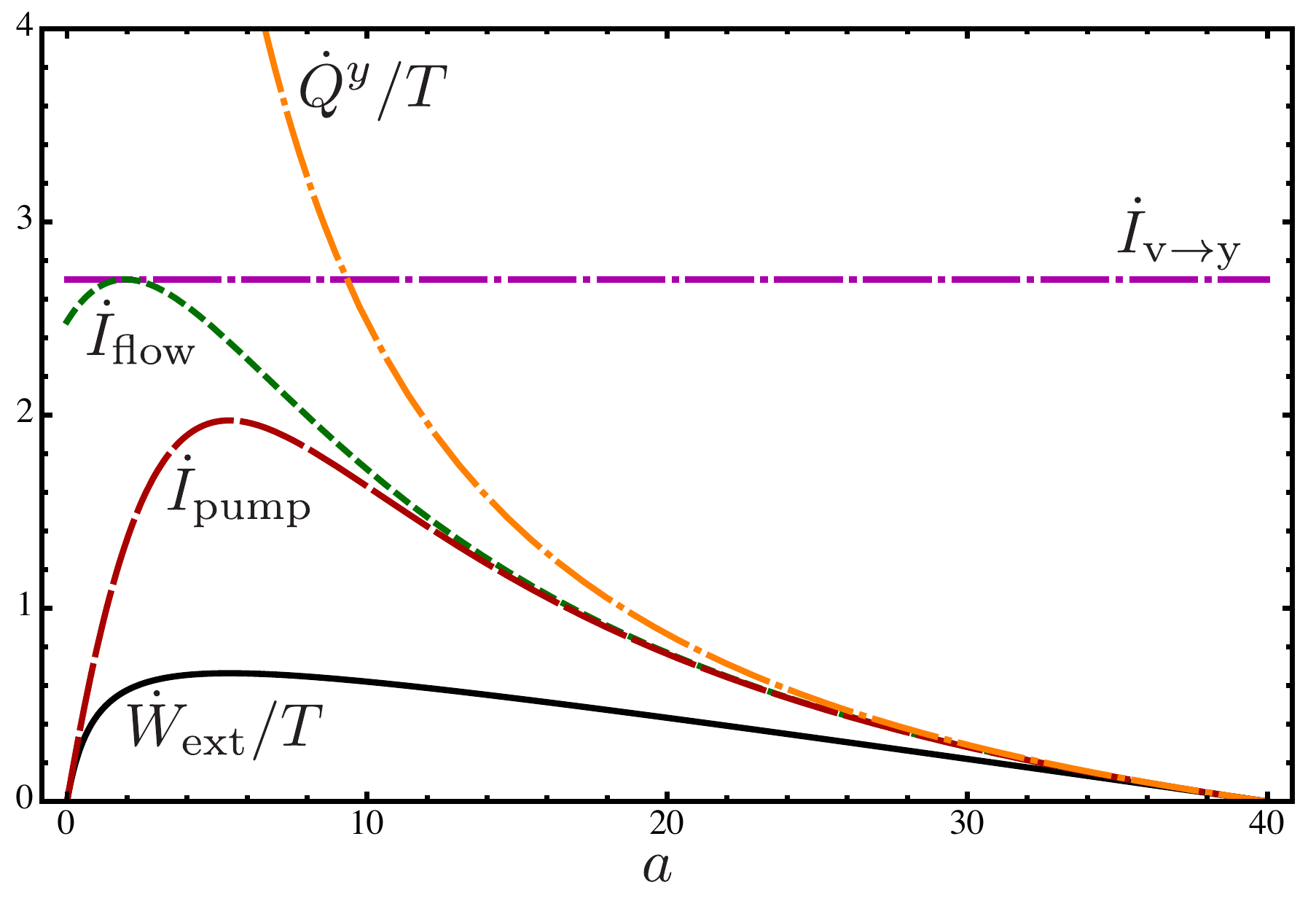}
\caption{Plot of  the heat dissipated by the auxiliary measurement oscillator ${\dot Q}^y$ as a function of the gain $a$ demonstrating that it upper bounds the information flow ${\dot I}_{\rm flow}$, entropy pumping ${\dot I}_{\rm pump}$, and extracted work ${\dot W}_{\rm ext}$. The transfer entropy rate ${\dot I}_{\rm v\to y}$ shares no relation with ${\dot Q}^y$. Parameters are $m=1$, $\gamma=1$, $T=5$, $\sigma=0.5$, and $\tau=0.1$.}
\label{fig:heat}
\end{figure}

When introducing the information flow, we divided the entropy production into two positive contributions \eref{eq:Ssplit}, one due  to the velocity ${\dot S}_{\rm i}^v$, and another due to the measurements ${\dot S}_{\rm i}^y$.
When studying the extracted work ${\dot W}_{\rm ext}$, we focused on ${\dot S}^v_{\rm i}$.
However, a similar analysis also holds for ${\dot S}^y_{\rm i}$, which verifies that the $y$-system must consume at least ${\dot I}_{\rm flow}$ free energy to sustain the correlations that promote feedback.
Observing that as a position $y_t$ is even under time-reversal (consistent with our previous analysis in \sref{sec:nofeedback}), we develop its thermodynamics by splitting its current $J^y_t$ \eref{eq:Curr} into irreversible and reversible portions as
\begin{equation}
\eqalign{
J^{{\rm irr, }y}_t(v,y)=-\frac{1}{\tau}yp_t(v,y)-\frac{T}{\tau}\partial_yp_t(v,y)\\
J^{{\rm rev, }y}_t(v,y)=\frac{v}{\tau}p_t(v,y)}.
\end{equation}
Notice that here the irreversible current is the time-symmetric contribution, since $y_t$ is even under time-reversal~\cite{Spinney2012}.
Then, repeating the analysis in \sref{sec:bipartite}, we have that in the steady state~\cite{Hartich2014,Allahverdyan2009,Horowitz2014}
\begin{equation}
{\dot S}^y_{\rm i}=\frac{\tau}{T}\int \frac{[J^{{\rm irr, }y}_{\rm s}(v,y)]^2}{p_{\rm s}(v,y)}\rmd v\rmd y=\frac{{\dot Q}^y}{T}-{\dot I}_{\rm flow}\ge 0,
\end{equation}
where ${\dot Q}^y=-\int y J^{\rm irr,y}_{\rm s}(v,y)\rmd v\rmd y=(\sigma^2_{\rm y}-T)/\tau$ is the heat flow rate into $y$'s reservoir.
Thus, in the steady state
\begin{equation}\label{eq:bipartiteHeat}
{\dot Q}^y\ge T{\dot I}_{\rm flow}.
\end{equation}
In order to track the velocity, $y$'s environment continually absorbs heat at a rate ${\dot Q}^y$.
We verify this bound in \fref{fig:heat}, where ${\dot Q}^y$ is plotted with ${\dot I}_{\rm flow}$.
The minimum ${\dot Q}^y=T{\dot I}_{\rm flow}$ is reached when the measurement device operates reversibly in the limit $\tau\ll \tau_{\rm v}$, so that $y_t$ rapidly relaxes to its instantaneous equilibrium centered about $v_t$: $p_{\rm s}(y|v)\propto \exp[-(y-v)^2/(2T)]$.

In addition, we have already argued that the entropy pumping lower bounds the  information flow, ${\dot I}_{\rm flow}\ge {\dot I}_{\rm pump}$ \eref{eq:biPuBounds}.
As a result, ${\dot I}_{\rm pump}$ offers a weaker lower bound on the energy required for an auxiliary system to provide the entropy-pumping feedback, ${\dot Q}^y\ge T{\dot I}_{\rm pump}$, which is verified in \fref{fig:heat} as well.

\subsection{Transfer entropy rate}\label{sec:transMeas}

The transfer entropy rate can also be understood as the minimum free energy required to measure, but with an alternative measurement scenario.
 In the previous section, the information flow was shown to bound the energy required to constantly rewrite a \emph{single} memory with each new measurement $y_t$.
 By contrast, the setup for the transfer entropy rate is  much closer to that envisioned by Landauer and Bennett in their thermodynamics of computation~\cite{Leff,Sagawa2010}: Each measurement is recorded separately in one of a  \emph{collection} of memories through a specific driven thermodynamic process; one example of which was recently described in \cite{Horowitz2013}.

The central idea is that each measurement outcome is recorded in a distinct memory.
Therefore, to track the system over any finite time interval, say from time $s=0$ to $t$, we require an infinite number of memories in which to record the infinity of measurements.
However, this is difficult to analyze.
So to proceed, we discretize time as $s_k=k\Delta s$, with $k=0,\cdots, N$ and $\Delta s=t/N$, where the measurement outcome at time $s_k$ is denoted simply as $y_{k}\equiv y_{s_k}$, and similarly $v_k\equiv v_{s_k}$.
To store these measurement outcomes, we imagine a collection of $N$ auxiliary memories with phase space positions $m_k$, prepared initially in positions $m_{k,0}$ distributed according to $\rho_0(m_{k,0})$.
The measurement is a thermodynamic process during a time interval of length $\theta$ in which the $k$-th memory is manipulated, with the velocity fixed, in such a way to reproduce  the correlations with $v_{k-1}$ embodied in the  measurement statistics of $y_k$.
In other words, we demand that the statistics of the $k$-th memory after the measurement are $m_{k,\theta}\sim y_k$ (equality in distribution).

\begin{figure}[htb]
\centering
\includegraphics[scale=.3]{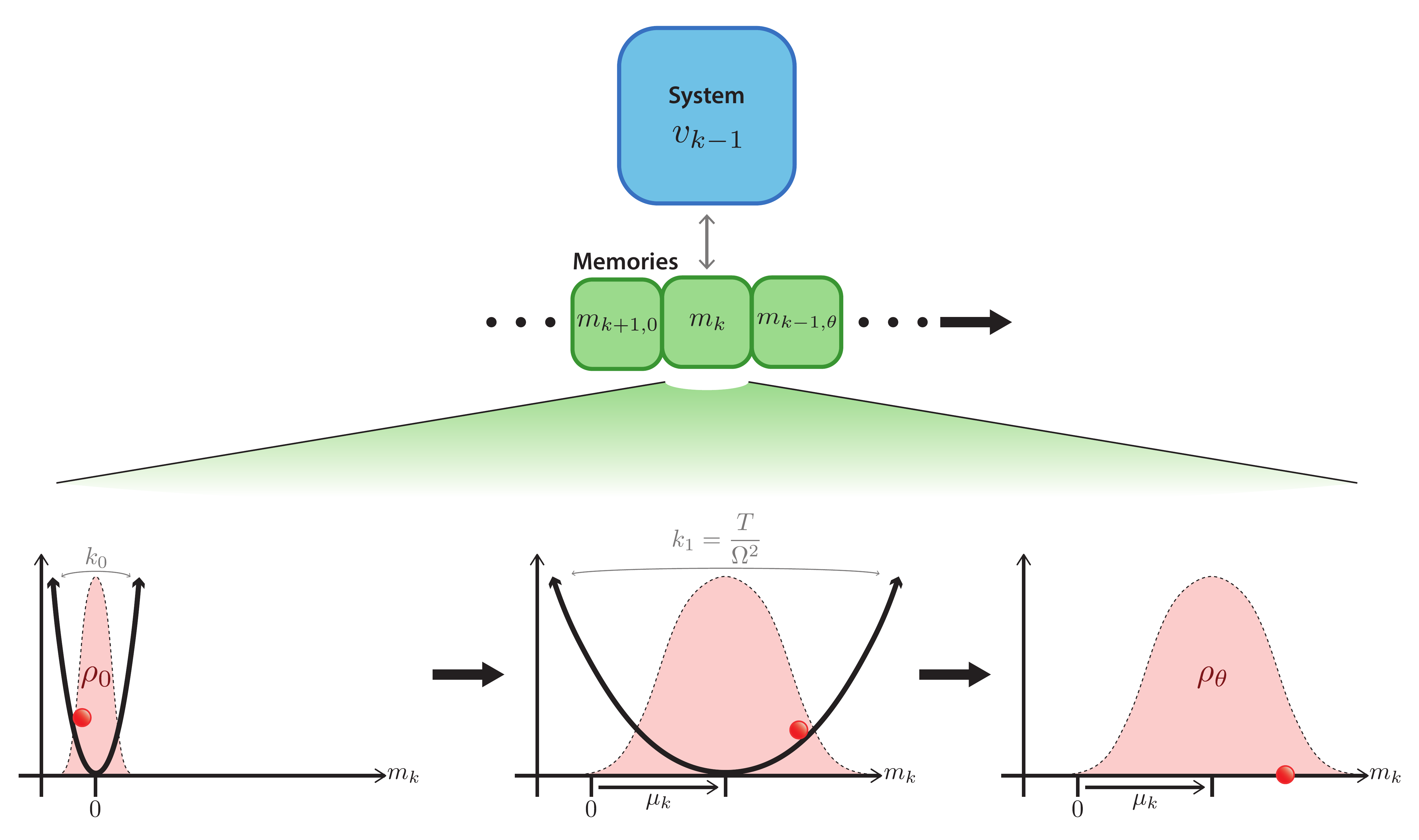}
\caption{Schematic illustration of the transfer entropy rate measurement scenario: At time $s_k$, the velocity $v_{k-1}$ is recorded in the the $k$-th memory, harmonic oscillator (red dot) with initial state $m_{k,0}$  through a nonautonomous interaction that slowly shifts and expands its potential $V(m_k,v_{k-1},m_{k-1})$, before quickly turning off. Concurrently, the probability density (pink shaded region) expands from $\rho_0(m_{k,0})$ to a width $\Omega^2$ and shifts by $\mu_k(v_{k-1},y_{k-1})$, terminating the process in the measurement probability density $\rho_\theta(m_{k,\theta})$ equivalent to \eref{eq:measProb}, correlated with $v_{k-1}$ \emph{and} the past measurement outcome $y_{k-1}$ stored in the previous memory state $m_{k-1,\theta}$.
The process is then repeated, with each new measurement recorded in the next memory the tape.}
\label{fig:tape}
\end{figure}

To see how these ideas play out in our model system, consider the discretized version of the Langevin equation \eref{eq:Langevin}
\begin{equation}\label{eq:LangevinDisc}
\eqalign{
y_{k}=y_{k-1}-\frac{\Delta s}{\tau}(y_{k-1}-v_{k-1})+\frac{1}{\tau}\Delta\eta_{k-1}},
\end{equation}
where the $\Delta\eta_k$ are independent Gaussian random variables of zero mean and covariance $\langle \Delta \eta_k\Delta\eta_l\rangle=\sigma^2\Delta s\delta_{kl}$.
\Eref{eq:LangevinDisc} is a rule that tells us how the measurement outcome $y_k$ at time $s_k$ depends on the velocity $v_{k-1}$ as well as the past measurement outcome $y_{k-1}$ stored in a previous memory.
Such measurements that depend on past outcomes are sometimes called non-Markovian measurements~\cite{Sagawa2011}.
Specifically,  $y_k$ is characterized by a Gaussian probability density
\begin{equation}\label{eq:measProb}
\eqalign{
P(y_{k}|y_{k-1},v_{k-1})=\frac{1}{\sqrt{2\pi\Omega^2}} \exp\left\{-\frac{[y_k-\mu_k(v_{k-1},y_{k-1})]^2}{2\Omega^2}\right\} \\
\mu_k(v_{k-1},y_{k-1})=y_{k-1}-\frac{\Delta s}{\tau}(y_{k-1}-v_{k-1}), \qquad \Omega^2=\frac{\sigma^2\Delta s}{\tau^2},
}
\end{equation}
roughly centered about the velocity with a variance depending on the measurement error.
Now, in view of our previous discussion, we desire a physical system to act as a memory and a measurement process that prepares that system in a statistical state with the probability density in \eref{eq:measProb}.
A natural choice is an overdamped harmonic oscillator coupled to a thermal reservoir at temperate $T$.
Initially each memory oscillator is prepared in equilibrium with an arbitrary initial spring constant $k_0$ centered about zero, as illustrated in \fref{fig:tape}.
Since each measurement is performed in sequence, it is attractive to visualize the phase spaces of the $N$ measurement oscillators aligned in a row, or tape.
Then one by one we couple each measurement oscillator to the system as well as past memories, so as to establish correlations.
The density in \eref{eq:measProb} suggests that the measurement protocol for the $k$-th oscillator should be the quasistatic turn-on of an interaction that shifts the center of the harmonic oscillator to $\mu_k$ -- which includes interactions with the past memories -- while simultaneously expanding the spring constant to $k_1=T/\Omega^2$, which results in the interaction potential
\begin{equation}
V(m_k,m_{k-1},v_{k-1})=\frac{T}{2\Omega^2}[m_k-\mu_k(v_{k-1},m_{k-1})]^2
\end{equation}
as depicted in \fref{fig:tape}.
As a result, upon completion of the $k$-th measurement the memory's position $m_{k,\theta}$ has settled into an equilibrium distribution $\rho_\theta(m_{k,\theta}|m_{k-1,\theta},v_{k-1})\propto \exp\left[-V(m_{k,\theta},m_{k-1,\theta},v_{k-1})/T\right]$ equivalent to~\eref{eq:measProb}.
To complete the measurement, we must freeze the state of the memory to lock in the correlations, and remove the interactions by turning off $V$.
One possible, though admittedly idealized, option is to instantaneously set $V=0$, and then immediately turn off the dynamics of the measurement oscillator -- perhaps by quenching the temperature to zero -- so that the oscillator can no longer move.
By repeating this sequence of actions on each new memory, we store a collection of measurement outcomes, each in a different physical memory.
Now to be precise each measurement has to be performed instantaneously from the point of view of the velocity.
This merely means that the time-scale of the evolution of the individual memories much be must faster than the velocity, $\theta\ll \tau_{\rm v}$, so that the measurement is completed before the velocity changes appreciably~\cite{Horowitz2013}.
However, this assumption is not unreasonable, since measurements are usually assumed to read out the instantaneous state of the system.

Having described how we can mimic the measurement statistics in a physical situation, we now address the thermodynamics from a general point of view, applying the methods of~\cite{Sagawa2012,Sagawa2013b,Horowitz2013}.
Our analysis is based on the following second-law-like inequality that relates the work performed in an isothermal process to the increment in the nonequilibrium free energy~\cite{Esposito2011,Deffner2012}: For a thermodynamic system with microscopic states $z$, the work $W$ performed along an isothermal process during which the system's probability density transitions from $p(z)$ to $p^\prime(z^\prime)$ is bounded as
 \begin{equation}\label{eq:Wbound}
 W-\Delta{\mathcal F}(z^\prime)\ge 0.
 \end{equation}
 where $\Delta {\mathcal F}(z^\prime)={\mathcal F}(z^\prime)-{\mathcal F}(z)$ is the change in the nonequilibrium free energy ${\mathcal F}(z)=U(z)-TS(z)$ defined in terms of the average internal energy $U(z)$ and entropy $S(z)=-\int p(z)\ln p(z)\, \rmd z$.
 The nonequilbirium free energy is a natural extension of the equilibrium free energy to systems characterized by an arbitrary probability density, since it reduces to the equilibrium free energy for systems in equilibrium.

 We begin by focusing on the work done during the $k$-th measurement, $W_k$, during which  the $k$-th memory becomes correlated with not only the velocity $v_{k-1}$ but all the past memories $m_0^{k-1}=\{m_{l,\theta}\}_{l=0}^{k-1}\sim\{y_l\}_{l=0}^{k-1}$ through the velocity which depends on the entire past.
Applying \eref{eq:Wbound}, we have
\begin{equation}\label{eq:Wk}
W_k-\Delta {\mathcal F}(m_{k,\theta}|m_0^{k-1},v_{k-1}) \ge 0,
\end{equation}
where $\Delta {\mathcal F}(m_{k,\theta}|m_0^{k-1},v_{k-1})={\mathcal F}(m_{k,\theta}|m_0^{k-1},v_{k-1})-{\mathcal F}(m_{k,0}|m_0^{k-1},v_{k-1})$ is the change in the nonequilibrium free energy of the $k$-th memory, corresponding to the change in the conditional density from $\rho_0(m_{k,0}|m_0^{k-1},v_{k-1})=\rho_0(m_{k,0})$ -- due to the initial independence of each memory -- to $\rho_\theta(m_{k,\theta}|m_0^{k-1},v_{k-1})=\rho_\theta(m_{k,\theta}|m_{k-1,\theta},v_{k-1})$.
We single out the new correlations by introducing the mutual information between $m_{k,\theta}$ and $v_{k-1}$ conditioned on all the past measurements as $I(m_{k,\theta};v_{k-1}|m_0^{k-1})=S(m_{k,\theta}|m_0^{k-1})-S(m_{k,\theta}|m_0^{k-1},v_{k-1})$~\cite{Cover}.
Substituting in this definition, \eref{eq:Wk} becomes
\begin{equation}
W_k-\Delta {\mathcal F}(m_{k,\theta}|m_0^{k-1}) \ge T I(m_{k,\theta};v_{k-1}|m_0^{k-1}),
\end{equation}
where $\Delta {\mathcal F}(m_{k,\theta}|m_0^{k-1})$ is the change in free energy conditioned on just the past memories: $\rho_0(m_{k,0})\to \rho_{\theta}(m_{k,\theta}|m_0^{k-1})$.
Summing over all measurements, we find
\begin{equation}\label{eq:transMeasDisc}
 W-\Delta {\mathcal F}(m_1^N|m_0)\ge TI^N_{\rm v\to y},
 \end{equation}
 where $W=\sum_{k=1}^N W_k$ is the work to perform all $N$ measurements, $\Delta {\mathcal F}(m_1^N|m_0)=\sum_{k=1}^N\Delta{\mathcal F}(m_{k,\theta}|m_0^{k-1})$ is the change in entire tape's free energy, and we have identified the discrete version of the transfer entropy~\cite{Ito2013},
\begin{equation}
  I^N_{\rm v\to y}=\sum_{k=1}^N I(m_{k,\theta};v_{k-1}|m_0^{k-1})
 \end{equation}
which is reviewed in \ref{sec:Disc}.
Importantly, by construction, the statistics of each memory reproduce the statistics of the measurement outcomes, so equivalently
\begin{equation}
 I^N_{\rm v\to y}=\sum_{k=1}^N I(y_k;v_{k-1}|y_0^{k-1})=\left\langle\ln\frac{{\hat \mathcal{P}}[y_0^N|v_0^N,y_0]}{{\mathcal P}[y_0^N|y_0]}\right\rangle.
\end{equation}
Taking the limit as the number of measurements go to infinity while the time between them goes to zero, we obtain
\begin{equation}\label{eq:transMeas}
{\dot W}-{\dot {\mathcal F}}= \lim_{t\to \infty}\frac{1}{t}\left[\lim_{\Delta s\to 0}W-\Delta {\mathcal F}(m_1^N|m_0)\right]\ge T{\dot I}_{\rm v\to y}.
\end{equation}
Thus, the transfer entropy rate is the minimum rate at which free energy is consumed to write to the memories.
The slow protocol that we described previously saturates this bound, since it is quasistatic and therefore thermodynamically reversible.

At this point, it is worthwhile to make a connection to a class of Maxwell-demon models that exploit a tape of low entropy, auxiliary systems or cells, similar to what we have just described~\cite{Barato2014,Mandal2012,Mandal2013,Deffner2013,Barato2013c,Hoppenau2014}.
Apart from the study in \cite{Hoppenau2014},  these models use an ideal tape that has no internal energy, and therefore cannot exchange energy with the system, but only entropy; a setup sometimes referred to as an information reservoir~\cite{Barato2014,Deffner2013}.
Under these conditions, a second-law-like inequality has been predicted that shows that the extracted work is bounded by the increase in entropy of each individual auxiliary system, ignoring the correlations between the different cells.
Our memories, on the other hand, have internal energy and therefore the natural thermodynamic quantity to consider is the free energy instead of  the entropy.
Therefore to fit our measurement model into this tape-model framework,
we must relate our information bound  on the work to measure to a bound that ignores the correlations.
To this end, we start with the bound for the energy to measure $W-\Delta {\mathcal F}(m_1^N|m_0)\ge TI^N_{\rm v\to y}$ in \eref{eq:transMeasDisc}, which includes through $\Delta{\mathcal F}$ all the correlations between different memories.
By noting that ignoring correlations and conditioning increases the entropy, $H(m_1^N|m_0)\le \sum_k H(m_{k,\theta})$~\cite{Cover}, we can conclude that ignoring the correlations decreases the free energy ${\mathcal F}(m_1^N|m_0)\ge \sum_k {\mathcal F}(m_{k,\theta})$.
As a result, we have from \eref{eq:transMeasDisc} and the initial independence of each memory the series of inequalities
\begin{equation}\label{eq:tapeModelBound}
W - \sum_{k=1}^N\Delta {\mathcal F}(m_{k,\theta})\ge T I^N_{\rm v\to y} \ge W_{\rm ext}.
\end{equation}
For the ideal tape with no internal energy this reduces to $\sum_k\Delta H(m_{k,\theta})\ge W_{\rm ext}$ recovering the ideal-tape bound~\cite{Barato2014,Mandal2012,Mandal2013,Deffner2013,Barato2013c,Hoppenau2014} in our setup.
\Eref{eq:tapeModelBound} may lead us to conclude that the bound on the extracted work from the tape-model framework, $W-\sum_{k=1}^N\Delta {\mathcal F}(m_{k,\theta})$, is weaker than that provided by the transfer entropy.
However, this would be too hasty, because these tape models allow a more general interaction between the tape cells and the system.
Whereas, in our setup the memory evolution is assumed to occur separately with the velocity fixed, the tape models consider a dynamics where the memory (or cell) would be allowed to evolve  simultaneously with the velocity.
From this point of view, the measurement model we have presented is a special case of these more general tape models, and it is exactly our assumption that the velocity is frozen during measurement that allows us to tighten the tape-model bound using the transfer entropy.  
Further comparisons of such tape models with other information measures and more traditional statements of the second law can be found in \cite{Horowitz2013,Barato2014}.

Finally, it should be noted that the preceding second law analysis can be viewed as a specific implementation of the information flow framework (outlined in sections \ref{sec:bipartite} and \ref{sec:bipartiteMeas}) applied to a nonautonomously driven auxiliary memory composed of a sequence of many subsystems, see \cite{Horowitz2014}.

\subsection{Discussion}

The transfer entropy rate and information flow both bound the energy consumed during  measurement.
However, each measurement scenario is distinct, and in general each of these information measures will not bound the energy consumption for the other's measurement scenario.
An example where ${\dot I}_{\rm v\to y}> {\dot Q}^y/T$ is possible is presented in \cite{Hartich2014};  thus, the transfer entropy rate does not generally lower bound  the amount of heat dissipated by a single memory being constantly rewritten.
Our model corroborates this observation, as verified in \fref{fig:heat} by the crossing of ${\dot Q}^y/T$ and ${\dot I}_{\rm v\to y}$.
The one exception is if the the controller implements the Kalman-Bucy filter \eref{eq:optcontrol}.
In which case, the equality of the information measures, ${\dot I}_{\rm flow}={\dot I}_{\rm v\to y}$, implies that there is a unique lower bound to the energy required for measurement.

To conclude this section, we take a broader perspective.
Our observation that the transfer entropy rate and  information flow both represent the minimum (free) energy consumed (or alternatively the entropy produced) in the auxiliary memory to create that information, suggests that it is reasonable to interpret some second-law-like inequalities as actually telling us something about the thermodynamics of the system \emph{and} its surroundings, where the surroundings include the measurement device.
 This allows us to incorporate information into the standard statement of the second law of thermodynamics through a kind of information reservoir on equal footing with other traditional thermodynamic reservoirs, similar to what was suggested for tapes in~\cite{Barato2014,Deffner2013}:
\begin{equation}
{\dot S}_{\rm i}={\dot S}+{\dot S}_{\rm env}={\dot S}-\frac{{\dot Q}}{T}+{\dot I}\ge 0,
\end{equation}
which is equivalent to \eref{eq:2law} in the steady state.
Here, ${\dot I}$ represents the minimum entropy produced in the environment that allows for feedback, with the minimum attained for reversible measurement.
The appropriate choice of ${\dot I}$ -- transfer entropy rate or information flow -- depends on which type of information reservoir we wish to use.
From this point of view, the efficiency $\varepsilon$ introduced in \eref{eq:eff} is a true measure of energetic efficiency that quantifies how faithfully the energy supplied by a reversible memory is extracted back out as work.


\section{Summary}

We have explored a collection of information measures that appear in second-law-like inequalities for  measurement and feedback, using the tools of stochastic thermodynamics and optimal control theory.
We have seen that these measures form a hierarchy of bounds on the extracted work, and that the Kalman-Bucy filter optimally will extract information and energy.
Even though each measure offers a different numerical bound on the extracted work, they also each correspond to different ways of gathering information.
With this distinction in mind, these second-law-like inequalities can be seen as manifestations of the second law of thermodynamics, since they include the entropy production of the system and surroundings, including the controller.


\ack We would like to thank Martin Rosinberg for a carefully reading of this manuscript. J.M.H. is supported financially by the ARO MURI grant W911NF-11-1-0268 and H.~S. is supported financially by the
Swedish Research Council under grant 2013-5523.


\appendix 

\section{Steady state probability density}\label{sec:ssDist}

The Gaussian steady state probability density in \eref{eq:ssDist} is completely characterized by its means, which are zero, and the covariance matrix $\bSigma$.
The elements of $\bSigma$ can be determined by exploiting the Fokker-Planck equation \eref{eq:FP} to develop a collection of equations for the variances $\langle v^2\rangle$, $\langle y^2\rangle$, and $\langle vy\rangle$, as described in \cite{Mazonka1999} for example.
The time-independent steady state solutions can then be shown to satisfy the algebraic equations
\begin{equation}
\eqalign{
\sigma_{\rm y}^2-\sigma_{\rm vy}=\frac{\sigma^2}{2\tau} \\
\frac{\gamma}{m}\sigma_{\rm v}^2+\frac{a}{m}\sigma_{\rm vy}=\frac{\gamma T}{m^2} \\
\left(\frac{\gamma}{m} +\frac{1}{\tau}\right)\sigma_{\rm vy}+\frac{a}{m}\sigma_{\rm y}^2-\frac{1}{\tau}\sigma_{\rm v}^2=0,
}
\end{equation}
whose solutions can be obtained after some lengthy algebra,
\begin{equation}
\fl \bSigma=
\left(\begin{array}{cc}
\frac{T}{m}\frac{1+(a/\gamma)(a\sigma^2/(2T))+(1+a/\gamma)(\gamma\tau/m)}{1+a/\gamma+(1+a/\gamma)(\gamma\tau/m)} & \frac{T}{m}\frac{1-a\sigma^2/(2T)}{(1+a/\gamma)(1+\gamma\tau/m)} \\
 \frac{T}{m}\frac{1-a\sigma^2/(2T)}{(1+a/\gamma)(1+\gamma\tau/m)}  & \frac{\sigma^2}{2\tau} \frac{1+a/\gamma+\gamma\tau/m+(2 T/m)/(\sigma^2/\tau)}{(1+a/\gamma)(1+\gamma\tau/m)}
\end{array}\right).
\end{equation}

\section{Path probabilities and the transfer entropy rate}\label{sec:Disc}

In this appendix, we demonstrate how we arrive at \eref{eq:P} for the trajectory probability density ${\mathcal P}$, and how this structure allows the compact expression for the transfer entropy rate in \eref{eq:trans}.

The analysis precedes by discretizing the evolution over the time interval $s=0$ to $t$ into steps of width $\Delta s=t/N$ as $s_k=k\Delta t$ for $k=0,\dots, N$ with $v_k\equiv v_{s_k}$ and $y_k\equiv y_{s_k}$.
We are interested in determining the probability density ${\mathcal P}[v_0^N,y_0^N]$ to observe the pair of discrete trajectories $v_0^N=\{v_k\}_{k=0}^N$ and $y_0^N=\{y_k\}_{k=0}^N$.
To this end, we discretize the Langevin equation \eref{eq:Langevin} as
\begin{equation}\label{eq:LangevinDisc2}
\eqalign{
mv_{k+1} & = mv_k-(\gamma v_k+ay_k)\Delta s+\Delta\xi_k\\
\tau y_{k+1} & = \tau y_k-(y_k-v_k)\Delta s + \Delta\eta_k}
\end{equation}
where $\Delta\xi_k$ ($\Delta\eta_k$) are independent, zero-mean, Gaussian random variable with covariance $\langle \Delta \xi_k\Delta\xi_l\rangle=2\gamma T\Delta s\delta_{kl}$ ($\langle \Delta \eta_k\Delta\eta_l\rangle=\sigma^2\Delta s\delta_{kl}$).
From this we deduce that to lowest order in $\Delta s$ the transition probability splits into separate $v$ and $y$ evolutions as~\cite{Allahverdyan2009}
\begin{equation}
\eqalign{
\fl P(v_{k+1},y_{k+1}|v_k,y_k)&=P(v_{k+1}|v_k,y_k)P(y_{k+1}|v_k,y_k) \\
&=\sqrt{\frac{m^2}{4\pi\gamma T\Delta s}}\exp\left[-\frac{(mv_{k+1}-mv_k+(\gamma v_k+ay_k)\Delta s)^2}{4\gamma T\Delta s}\right]\\
&\hspace*{2cm}\times\sqrt{\frac{\tau^2}{2\pi \sigma^2\Delta s}}\exp\left[-\frac{(\tau y_{k+1}-\tau y_k-(y_k-v_k)\Delta s)^2}{2\sigma^2\Delta s}\right].
}
\end{equation}
Thus, the joint trajectory probability takes the form
\begin{equation}
\fl {\mathcal P}[v_0^N,y_0^N]={ P}(v_{N}|v_{N-1},y_{N-1}){ P}(y_{N}|v_{N-1},y_{N-1})\cdots { P}(v_{1}|v_0,y_0){ P}(y_{1}|v_0,y_0)p(v_0,y_0),
\end{equation}
with arbitrary initial density $p(v_0,y_0)$.
Since the evolution naturally divides, it suggests introducing the trajectory conditional probabilities
\begin{eqnarray}
\hat{\mathcal P}[v_0^N|y_0^N,v_0]&={ P}(v_{N}|v_{N-1},y_{N-1})\cdots { P}(v_{2}|v_1,y_1){ P}(v_{1}|v_0,y_0)\\
\hat{\mathcal P}[y_0^N|v_0^N,y_0]&={ P}(y_{N}|v_{N-1},y_{N-1})\cdots { P}(y_{2}|v_1,y_1){ P}(y_{1}|v_0,y_0),
\end{eqnarray}
in terms of which the joint trajectory probability becomes
\begin{equation}
{\mathcal P}[v_0^N,y_0^N]=\hat{\mathcal P}[v_0^N|y_0^N,v_0]\hat{\mathcal P}[y_0^N|v_0^N,y_0]p(v_0,y_0).
\end{equation}
Equations \eref{eq:Phat1}, \eref{eq:Phat2}, and \eref{eq:P} are the continuous time versions of the preceding equations obtained in the limit $\Delta s\to 0$.

In this discretized setup, we can directly apply the theory of discrete feedback~\cite{Sagawa2008,Horowitz2010, Ponmurugan2010,Hartich2014,Suzuki2009,Barato2013b}.
Here, the transfer entropy after $N$ measurements is given as
\begin{eqnarray}\label{eq:transDisc}
 I_{\rm v\to y}^N &=\sum_{k=0}^{N-1}\int  P(v_k,y_0^{k+1})\ln \left[\frac{P(y_{k+1}|v_k,y_k)}{P(y_{k+1}|y_0^k)}\right] \rmd v_k \rmd y_0^{k+1}.
\end{eqnarray}
We see that the transfer entropy is the relative entropy between the transition probability of $y$ given $v$, $P(y_{k+1}|v_k,y_k)$,  and the unconditioned transition probability, $P(y_{k+1}|y_0^k)$, averaged over $(v_k,y_0^k)$.
Recall that the relative entropy between two probability densities $f(x)$ and $g(x)$ is $D(f||g)=\int f(x)\ln [f(x)/g(x)]\rmd x$~\cite{Cover}.
In this way, the transfer entropy measures the affect the velocity has on the measurement dynamics, that is, how distinguishable the measurement dynamics given the velocity are from the measurement dynamics without the velocity.
Expanding the sum we can rewrite \eref{eq:transDisc} as
\begin{equation}
I_{\rm v\to y}^N=\int \rmd v_0^N\rmd y_0^N {\mathcal P}[v_0^N,y_0^N]\ln \frac{\hat{\mathcal P}[y_0^N|v_0^N,y_0]}{{\mathcal P}[y_0^N|y_0]}.
\end{equation}
The continuous time version appears in \eref{eq:trans}.

\section{Power spectra formulae for information rates}\label{sec:power}

In this appendix, we sketch how entropy rates for stationary Gaussian processes can be expressed in terms of the processes' correlation functions, following the developments in~\cite{Munakata2006,Tostevin2010}.

Let us consider a discretization with spacing $\Delta s=t/N$ of a Gaussian stochastic process ${\vec x}=\{x_k\}_{k=0}^N$.
It is completely characterized by its mean $\vec{\mu}=\{\mu_k\}=\{\langle x_k\rangle\}$ and covariance matrix ${\bf C}$ with elements $C_{mn}=\langle (x_m-\mu_m)(x_n-\mu_n)\rangle$, which we assume to be time-independent, $C_{mn} = c(|m-n|)$, an example being a stationary process:
\begin{equation}
{\mathcal P}({\vec x})=\frac{1}{\sqrt{(2\pi)^N|{\bf C}|}}\exp\left[-\frac{1}{2}({\vec x}-\vec\mu)\cdot {\bf C}^{-1}\cdot({\vec x}-\vec\mu)\right].
\end{equation}

The power spectra formulae for the information rates  follow from the observation that the entropy of such a Gaussian distribution is completely characterized by the covariance matrix:
\begin{equation}
H({\vec x})=\frac{N}{2}\ln (2\pi\rme) +\frac{1}{2}\ln|{\bf C}|.
\end{equation}

Since the process is causal, the covariance matrix has a Toeplitz structure, $C_{mn}=c(|m-n|)$, which allows us to diagonalize it in the limit $N\to\infty$ using its Fourier transform $C(\omega)=\sum_{s=0}^N e^{-i\omega s}c(s)$, with $\omega=2\pi/t$.
In which case, the entropy rate can be expressed as~\cite{Tostevin2010}
\begin{equation}\label{eq:entRate1}
 {\dot H}=\lim_{N\to \infty}\frac{1}{\Delta s  N}H({\vec x})=\frac{1}{2\Delta s}\ln (2\pi\rme)+\frac{1}{2}\int_{-\pi/\Delta s}^{\pi/\Delta s} \ln C(\omega)\rmd \omega.
\end{equation}

The transfer entropy is the difference in entropy rate between the trajectory of measurement outcomes ${\mathcal P}[y_0^N|y_0]$ and the entropy rate for $\hat{\mathcal P}[y_0^N|v_0^N,y_0]$:
\begin{equation}
\fl {\dot I}_{\rm v\to y}=\lim_{N\to\infty}\frac{1}{\Delta sN} \left[H(y_0^N|y_0)-{\hat H}(y_0^N|v_0^N,y_0)\right]=-\frac{1}{4\pi}\int_{-\pi/\Delta s}^{\pi/\Delta s}\ln\frac{{\hat C}_{\rm yy|v}(\omega)}{C_{\rm yy}(\omega)} \rmd\omega.
\end{equation}
Taking the continuous time limit $\Delta s\to0$, we recover the expression in \eref{eq:trans}.
Similarly, the trajectory mutual information is
\begin{equation}\label{eq:trajDisc}
\eqalign{
{\dot I}_{\rm traj}&=\lim_{N\to\infty}\frac{1}{\Delta sN}\left[H(y_0^N)+{ H}(v_0^N)-{ H}(v_0^N,y_0^N)\right] \\
&=-\frac{1}{4\pi}\int_{-\pi/\Delta s}^{\pi/\Delta s}\ln\frac{{\mathcal C}(\omega)}{C_{\rm vv}(\omega)C_{\rm yy}(\omega)} \rmd\omega,
}
\end{equation}
where ${\mathcal C}(\omega)$ is the Fourier transform of the covariance matrix of the joint measurement and velocity process.
One can show, as in \cite{Munakata2006}, that
\begin{equation}
{\mathcal C}(\omega)=C_{\rm vv}(\omega)C_{\rm yy}(\omega)-|C_{\rm vy}(\omega)|^2,
\end{equation}
which when substituted into \eref{eq:trajDisc} recovers \eref{eq:trajOmega} after the taking $\Delta s\to0$.

\section{Calculation of information rates}\label{sec:Integral}

In this appendix we calculate ${\dot I}_{\rm v\to y}$ in \eref{eq:transOmega} and ${\dot I}_{\rm traj}$ in \eref{eq:trajOmega}.
As a first step, we must determine the Fourier transforms of  various correlation functions.
To this end, we begin by Fourier transforming the equations of motion for $v_t$ and $y_t$ in \eref{eq:Langevin}:
\begin{equation}\label{eq:LangevinOmega}
\eqalign{ im\omega \hat v_\omega=-\gamma \hat v_\omega-a\hat y_\omega+\hat\xi_\omega\\
i\tau\omega \hat y_\omega=-(\hat y_\omega-\hat v_\omega-\hat \eta_\omega)},
\end{equation}
with $\langle |\hat \xi _\omega|^2\rangle=2\gamma T$ and $\langle |\hat \eta _\omega|^2\rangle=\sigma^2$.

Let us start by determining ${\dot I}_{\rm v\to y}$, which requires two correlation functions obtained from the solutions of \eref{eq:LangevinOmega} as
\begin{equation}
{\hat C}_{\rm yy|v}(\omega)=\left\langle\left|{\hat y}_\omega-\frac{{\hat v}_\omega}{i\tau\omega+1}\right|^2\right\rangle=\frac{\sigma^2}{\tau^2\omega^2+1},
\end{equation}
and
\begin{equation}
\eqalign{
C_{\rm yy}(\omega)=\langle |{\hat y}_\omega|^2\rangle&=\left|\frac{1}{1+\frac{a}{(i\tau\omega+1)(im\omega+\gamma)}}\right|^2\frac{1}{\tau^2\omega^2+1}\left(\frac{2\gamma T}{m^2\omega^2+\gamma^2}+\sigma^2\right)\\
&\equiv|{\mathcal S}|^2\frac{1}{\tau^2\omega^2+1}\left(\frac{2\gamma T}{m^2\omega^2+\gamma^2}+\sigma^2\right),
}
\end{equation}
where $\mathcal S$ is known as the sensitivity function of the feedback system \cite{Astrom}.
Thus, the transfer entropy rate is
\begin{equation}
{\dot I}_{\rm v\to y}=-\frac{1}{4\pi}\int_{-\infty}^\infty\underbrace{\ln\left(\frac{\sigma^2(m^2\omega^2+\gamma^2)}{\sigma^2(m^2\omega^2+\gamma^2)+2\gamma T}\right)}_A+\underbrace{\ln|{\mathcal S}|^2}_B\rmd \omega.
\end{equation}
These integrals can be performed by exploiting the formula~\cite{Gradshteyn}
\begin{equation}\label{eq:int}
\int_{0}^\infty \ln\left(\frac{z^2+a^2}{z^2+b^2}\right)\rmd z=\pi(a-b).
\end{equation}
In particular,
\begin{equation}
\fl A=-\frac{1}{2\pi}\int_{0}^\infty\ln\left(\frac{\omega^2+(\gamma/m)^2}{\omega^2+(\gamma/m)^2+2\gamma T/(m^2\sigma^2)}\right)\rmd\omega=\frac{\gamma}{2m}\left(\sqrt{1+\frac{2T/\gamma}{\sigma^2}}-1\right),
\end{equation}
and
\begin{equation}
B=-\frac{1}{4\pi}\int_{-\infty}^\infty\ln|{\mathcal S}|^2\rmd \omega=0,
\end{equation}
which recovers \eref{eq:transCalc}. 
That logarithmic integrals of the sensitivity function, such as $B$, equals zero holds
with great generality. In fact, it represents a well-known conservation principle in control theory 
known as Bode's integral formula~\cite{Astrom}.

To determine ${\dot I}_{\rm traj}$, we first note that ${\dot I}_{\rm traj}={\dot I}_{\rm v\to y}+{\dot I}_{\rm y\to v}$.
Since we already know ${\dot I}_{\rm v\to y}$, it remains to determine
\begin{equation}
{\dot I}_{\rm y\to v}=-\frac{1}{4\pi}\int_{-\infty}^{\infty}\ln\frac{{\hat C}_{\rm vv|y}(\omega)}{C_{\rm vv}(\omega)}\rmd\omega.
\end{equation}
The power spectra are obtained from \eref{eq:LangevinOmega} as
\begin{equation}
\eqalign{
{\hat C}_{\rm vv|y}(\omega)=\left\langle\left|{\hat v}_{\omega}+\frac{a{\hat y}_\omega}{im\omega+\gamma}\right|^2\right\rangle=\frac{2\gamma T}{m^2\omega^2+\gamma^2}\\
C_{\rm vv}(\omega)=\langle|{\hat v}_\omega|^2\rangle=|{\mathcal S}|^2\frac{1}{m^2\omega^2+\gamma^2}\left(\frac{a^2\sigma^2}{\tau^2\omega^2+1}+2\gamma T\right).
}
\end{equation}
Therefore, recognizing that the contribution from the sensitivity function ${\mathcal S}$ is zero, we have
\begin{equation}
\fl {\dot I}_{\rm y \to v}=-\frac{1}{4\pi}\int_{-\infty}^\infty\ln\left(\frac{\omega^2+1/\tau^2}{\omega^2+1/\tau^2+a^2\sigma^2/(2\gamma T \tau^2)}\right)=\frac{1}{2\tau}\left(\sqrt{1+\frac{a^2\sigma^2}{2\gamma T}}-1\right),
\end{equation}
by virtue of \eref{eq:int}.


\section*{References}

\bibliography{Feedback.bib,PhysicsTexts.bib}
\bibliographystyle{iopart-num}
\end{document}